\documentclass[%
aps,
pre,preprint,
superscriptaddress
]{revtex4-1}

\usepackage{graphicx}% Include figure files
\usepackage{dcolumn}% Align table columns on decimal point
\usepackage{bm}% bold math
\begin{document}

%\preprint{}

%%%%%%%%%%%%%%%%%%%%%%%%%%%%%%%%%%%%%%%%%%%%%%%%%%%%%%%%%%%%%%%%%
%\title{The Dynamic Mechanism of a 2D Moving Crookes Radiometer}% Force line breaks with \\

%\affiliation{%


%\affiliation{
%\affiliation{
 %Third institution, the second for Charlie Author
%\affiliation{%

% Prandtl number for non-equilibrium gas flow
\title{A simplification of the unified gas kinetic scheme}
\author{Songze Chen}
\email{jacksongze@hust.edu.cn}
\affiliation{
State Key Laboratory of Coal Combustion, Huazhong University of Science and technology, Wuhan 430074, China
}%
\author{Zhaoli Guo}
\email{zlguo@hust.edu.cn}
\affiliation{
State Key Laboratory of Coal Combustion, Huazhong University of Science and technology, Wuhan 430074, China
}%
\author{Kun Xu}%
\email{makxu@ust.hk}
\affiliation{
The Hong Kong University of Science and technology Clear Water Bay, Kowloon, Hong Kong, China
}%

\date{\today}% It is always \today, today,

\begin{abstract}
Unified gas kinetic scheme (UGKS) is an asymptotic preserving scheme for the kinetic equations. It is superior for transition flow simulations, and has been validated in the past years.  However, compared to the well known discrete ordinate method (DOM) which is a classical numerical method solving the kinetic equations, the UGKS needs more computational resources. In this study, we propose a simplification of the unified gas kinetic scheme. It allows almost identical numerical cost as the DOM, but predicts numerical results as accurate as the UGKS. Based on the observation that the equilibrium part of the UGKS fluxes can be evaluated analytically, the equilibrium part in the UGKS flux is not necessary to be discretized in velocity space. In the simplified scheme, the numerical flux for the velocity distribution function and the numerical flux for the macroscopic conservative quantities are evaluated separately. The simplification is equivalent to a flux hybridization of the gas kinetic scheme for the Navier-Stokes (NS) equations and conventional discrete ordinate method. Several simplification strategies are tested, through which we can identify the key ingredient of the Navier-Stokes asymptotic preserving property. Numerical tests show that, as long as the collision effect is built into the macroscopic numerical flux, the numerical scheme is Navier-Stokes asymptotic preserving, regardless the accuracy of the microscopic numerical flux for the velocity distribution function.

%\begin{description}
%\end{description}

%\keywords{Asymptotic preserving \sep Unified gas kinetic schemes}%Use showkeys class option if keyword
%                              %display desired
\end{abstract}

\maketitle

%\tableofcontents

%\paragraph{Syntax}
%\paragraph{Syntax}
\section{Introduction}
In recent years, multiscale computation is recognized as a powerful tool for studying the interaction on different scales and/or different hierarchies. It has become an active research field and has been applied in many areas, for instance, rarefied gas dynamics, radioactive, plasma, and phonon transfer.

In the rarefied gas dynamics, the physical scales are characterized by the typical geometric length ($L$) and mean free path ($\lambda$). % which is defined as the mean distance that a particle travels between two successive particle collisions.
The ratio of these two characteristic lengths is known as the Knudsen number (Kn$=\lambda/L$).
When the Knudsen number is much smaller than 1, it is well known that the Navier-Stokes equations are established and govern the fluid behavior. But when the Knudsen number is not too small, the Navier-Stokes equations do not provide accurate physics, and the kinetic equation should be adopted as the governing equation. The simplest kinetic equation for monatomic gas is the BGK equation \cite{BGK1954}, which takes the following form,
\begin{eqnarray}
  \frac{\partial f}{\partial t} + \mathbf{u}\cdot \nabla f = \frac{g-f}{\tau}, \label{eq:BGK}
\end{eqnarray}
where $f$ represents the particle velocity distribution function depending on the location ($\mathbf{x}$), the time ($t$), and particle velocity ($\mathbf{u}$), $g$ denotes the corresponding equilibrium state shown as follows,
\begin{eqnarray}
g &=& \mathcal{G}(W) = \rho\left\{\frac{2RT}{\pi}\right\}^{3/2}\exp(-\frac{1}{2RT}(\mathbf{u-U})^2),\label{eq:equilibrium}\\
W &=& (\rho, \rho \mathbf{U}, \rho E)^T.  \label{eq:macroVar}
\end{eqnarray}
where $\rho$ is the gas density, $T$ is the gas temperature, $R$ is the gas constant, and $E$ denotes the total energy. Since the collision process is conserved, $g$ and $f$ share identical conservative quantities, that is,
\begin{eqnarray}
\langle \psi g \rangle = \langle \psi f \rangle, \ \psi = (1, \mathbf{u}, \frac{1}{2}\mathbf{u}^2)^T.
\end{eqnarray}
The symbol $\langle f \rangle$ is defined as, $\langle f \rangle = \int_{-\infty}^{+\infty} f d\mathbf{u}$.

Typically, the flow regimes can be categorized into four regimes: continuum flow ($\mbox{Kn} < 0.001$), slip flow ($0.001 < \mbox{Kn} < 0.1$), transition flow ($0.1 < \mbox{Kn} < 10$), and free molecular flow ($\mbox{Kn} > 10$). The Navier-Stokes equations are only validated in the continuum flow regime, and can be further extended to solve a small portion of slip flow problems by considering slip boundary condition. For the other flow regimes, the kinetic theory, including Boltzmann equation and other kinetic equations, must be adopted to take account of the delicate molecular motion. For example, when a vehicle travels through the atmosphere, the density of ambient gas changes dramatically. In another scenario, the gas is driven by the temperature gradient, goes through different chambers in multistage Knudsen pump. The mean free path enlarges as the density decreases, and the Knudsen number enlarges accordingly. The Navier-Stokes equations fail to predict the flow fields somewhere in these two applications. Thus the kinetic equation is necessary to take over in the domain where NS equations break down.
An intuitive idea is the domain decomposition method, in which the flow field is solved on different subdomains by appropriate numerical solvers, the Navier-Stokes solvers or the kinetic solvers. But the major difficulty of this method is the information exchange in the buffer zone or overlap region between two numerical methods on different scales. Moreover, in many multiscale problems, the Knudsen number varies both in space and time. Single domain decomposition is incapable for such problems.

Another promising multiscale approach is the asymptotic preserving scheme that can recover large scale system from small scale simulation uniformly\cite{Jin2012}. When the Knudsen number goes to zero, the numerical scheme for the kinetic equation should be an analogue of the analytical asymptotic analysis of the kinetic equation. In 1991, Coron and Perthame \cite{Coron1991} proposed a scheme which is asymptotic preserving in terms of Euler equations. After this study, variants AP schemes for the rarefied gas system are proposed in the last two decades, including implicit scheme for the collision terms \cite{Pieraccini2007,Filbet2011}, penalization method \cite{Gabetta1997, Filbet2010, Yan2013}, exponential relaxation method \cite{Coron1991,Dimarco2011}, unified gas kinetic schemes \cite{Xu2010,ugks1_1,ugks2,ugks3}, and discrete unified gas kinetic scheme \cite{guo2013discrete,guo2014discrete,wang2015comparative} etc.

From the previous literatures, two key ingredients of the asymptotic preserving scheme can be concluded.
The first key ingredient is the special treatment of the collision term (RHS of Eq.(\ref{eq:BGK})).
The traditional DOM solves the collision term explicitly. It is always restricted by the Knudsen number, and cannot obtain physical solution in near continuum and continuum flow regime unless using infinite computation resources. Actually, the stiffness of the collision term due to the small parameter makes the explicit schemes for the kinetic equation useless in the continuum flow regime.
Therefore, the exponential collision solver \cite{Coron1991, Dimarco2011, li2014exponential} and implicit treatment of the collision term \cite{Gabetta1997,Filbet2010,Caflisch1997} are proposed to remove the stiffness of the collision term.

%clarify the terminology, the numerical flux and the numerical body force.

The other ingredient of the AP scheme is that the completed kinetic equation must be employed to solve the numerical flux at cell interface and the body force inside a cell in order to attain the correct Navier-Stokes limit \cite{chen2015comparative}. Bennoune \emph{et al}.\cite{Bennoune2008} investigated the influence of the implicit schemes for the collision term, and found that, if operator splitting method is employed to evaluate the collision term, the resulting distribution function will be too close to the equilibrium state, thus the schemes cannot attain the physical viscosity. Chen and Xu\cite{chen2015comparative} studied the Navier-Stokes asymptotic preserving property and concluded that not only the body force needs both convection and collision terms, the numerical fluxes also need these two terms in order to obtain the correct Navier-Stokes limit.

In the early stage, the operator splitting method is employed to simplify the numerical scheme.
The governing equation is modified for different purposes. For solving the interfacial numerical flux, the convection term is reserved, but the source term is discarded. Governing equation becomes,
\begin{eqnarray}
  \frac{\partial f}{\partial t} + \mathbf{u}\cdot \nabla f = 0. \label{eq:collisionless}
\end{eqnarray}
For solving the body force, only the source term is reserved, while convection term is abandoned.
\begin{eqnarray}
  \frac{\partial f}{\partial t} = \frac{g-f}{\tau}. \label{eq:onlyCollision}
\end{eqnarray}
It is found that the use of incomplete governing equation will induce large error when simulate continuum flows \cite{chen2015comparative, Bennoune2008}.

In 2010, Xu \emph{et al} proposed the unified gas kinetic scheme, which couples the collision and convection terms by a local analytical solution of the complete governing equation (\ref{eq:BGK}). When approaching the Navier-Stokes limit, the numerical flux turns to the Chapman-Enskog expansion gradually. Therefore, the collision and free transport are all built into the numerical flux and the numerical body force. Theoretically, UGKS can recover the NS limit and Euler limit.
With the same spirit, Guo \emph{et al.} proposed a discrete unified gas kinetic scheme (DUGKS) which replaces the local integral solution by a discrete time integral. Although, the discrete approximation is adopted, the DUGKS still possess the NS AP property.

Theoretically, the unified schemes can recover the continuum regime. However, quadrature which accounts for the numerical integral in discrete velocity space is an obstacle for attaining correct asymptotic limit in the continuum flow regime.
%1. The low order accuracy of collision term. Since the initial Chapman-Enskog expansion for each time step is derived by the collision term at last time step, there is always a first order error $O(\tau\Delta t)$ in the viscous term as long as the explicit discretization is employed to calculate the numerical fluxes.
As we know, in the free molecular flow regime, the Newton-Cotes quadrature is more suitable compared to the Gauss-Hermite quadrature because the distribution function deviates largely from equilibrium state. But in the continuum flow, the Gauss-Hermite quadrature is always used due to its high accuracy for the integral of exponential function. If different quadratures are employed, massive interpolations will be needed to exchange data on different velocity points. And it will introduce additional numerical errors. As a result, it is inconvenient to change the quadrature method automatically according to the flow condition. Therefore, unsuitable quadrature might induce large error or large computational cost in a unified AP scheme.

In practice, a unified scheme is still burdensome to reproduce the continuum flow limit. On the other hand, the Navier-Stokes equations can be derived from the kinetic equation, and the traditional numerical schemes for the Navier-Stokes equations are highly efficient. Why do we derive asymptotic limit from massive high dimensional distribution function in numerical scheme? If we use more degree of freedom to simulate a lower dimensional problem, then there must be something can be simplified. In this study, we revisit the unified gas kinetic scheme and estimate the contribution of each term in asymptotic limit. For the part which can be calculated by traditional Navier-Stokes solver, we use analytical results instead of the discrete velocity representation and propose several simplification of the UGKS.

The article is organized as follows. In Sec. 2, the unified gas kinetic scheme is introduced briefly; in Sec. 3, we analyze the behavior of the UGKS in different flow regime and propose three different simplification strategies; in Sec. 4, the numerical discretization and the boundary condition are introduced;
in Sec. 5, numerical comparisons are provided, from which the key ingredient of the unified scheme and the best simplification strategy are identified for the industrial applications. Finally, we conclude this study in Sec. 6.

%\begin{itemize}
%\item About terminology
%\end{itemize}
\textbf{Remark:} We shall emphasize the terminologies used above. As the operator splitting method has been prevailing for many years, the numerical flux is always correlated with the convection term (Eq.(\ref{eq:collisionless})), and the body force is correlated with the collision term (Eq.(\ref{eq:onlyCollision})). However, during a finite time interval, the interfacial flux is not only influenced by the convection term, but is also influenced by collision term, and so is the body force. In this paper, we do not use the 'convection' and 'collision' to illustrate the two procedures in the numerical scheme.
Actually, considering a control volume, the quantities changing inside the control volume equals to the interfacial flux through the interface plus the body force exerted on the volume. The interfacial flux and the body force are only geometric concepts in the finite volume schemes. Thus, the terminologies, 'interfacial flux' and 'body force', are precise to describe the two procedures in the unified schemes.

\section{Unified gas kinetic scheme}
In this paper, we only consider the finite volume schemes. We will fix the numerical method for the body force, and compare different interfacial fluxes.
Before discussing the AP property of the UGKS, we briefly recall the numerical flux of the conventional DOM for the kinetic equation.
As mentioned in the introduction, the collisionless kinetic equation (Eq.(\ref{eq:collisionless})) is taken as the governing equation to evaluate the numerical flux.
The solution at the interface ($\mathbf{x}=0$) is then,
\begin{eqnarray}
    f(0,t,\mathbf{u}) = f(-t\mathbf{u},0,\mathbf{u}).
\end{eqnarray}
Considering first order spatial expansion, we have,
\begin{eqnarray}
    f_{dom}(0,t,\mathbf{u}) = f(0,0,\mathbf{u}) - t\mathbf{u}\cdot \nabla f.
\end{eqnarray}
The numerical flux for the distribution function is then,
\begin{eqnarray}
    \mathcal{F}_{dom} &=& \int_{0}^{\Delta t} u_k f_{dom}(0,t,\mathbf{u}_k) dt = u_k (\Delta t f_k-\frac{1}{2}\Delta t^2 \mathbf{u}_k\cdot \nabla f_k). \label{eq:domflux}
\end{eqnarray}
For simplicity, we ignore the arguments of the distribution function $f$ and assume that $u$ is aligned with the normal direction of the cell interface. The numerical flux of the DOM is very simple, only the numerical fluxes for the distribution function are considered in the DOM. In order to compare with the UGKS, the equivalent numerical fluxes for macroscopic variables are derived by taking the moments of the numerical microscopic flux,
%\mathcal{F}^W_{dom} &=& \sum_k u_k \psi_k (\Delta t f_k-\frac{1}{2}\Delta t^2 \mathbf{u}_k\cdot \nabla f_k),
\begin{eqnarray}
    \mathcal{F}^W_{dom} &=& \langle \mathcal{F}_{dom} \rangle_k = \sum_k u_k \psi_k (\Delta t f_k-\frac{1}{2}\Delta t^2 \mathbf{u}_k\cdot \nabla f_k), \label{eq:domfluxW}
\end{eqnarray}
where the superscript $W$ denotes the macroscopic flux. The symbol $\langle f \rangle_k$ denotes taking moments of $f$ in discrete velocity space, namely, the summation $\langle f \rangle_k = \sum_k \omega_k f_k$, where $\omega_k$ is the weight function at velocity point $\mathbf{u}_k$. The mechanism of the above formulations for the macroscopic fluxes is equivalent to kinetic flux vector splitting (KFVS) method for the Euler equations.

The unified gas kinetic scheme is an asymptotic preserving scheme benefiting from the local analytical solution of kinetic equation.
Integrating along the characteristic of the BGK equation (Eq.(\ref{eq:BGK})), a local analytical solution can be derived.
\begin{eqnarray}
  f(0,t,\mathbf{u}) &=& e^{-t/\tau}f(-\mathbf{u}t,0,\mathbf{u}) \nonumber\\
   && + \frac{1}{\tau}\int_0^{t}g(-\mathbf{u}(t-t'),t', \mathbf{u})e^{-(t-t')/\tau}dt'. \label{eq:localsolution}
\end{eqnarray}
The forepart is the non-equilibrium part. When the system approaches equilibrium, $e^{-t/\tau}$ will become zero asymptotically, i.e., the non-equilibrium contribution will vanish.
Meanwhile, the second term on the right hand side, which represents the equilibrium part, will dominate.

Suppose, after the numerical reconstruction, the physical quantities are linearly distributed around the cell, and are expressed as follows,
\begin{eqnarray}
    f(\mathbf{x},0,\mathbf{u}) &=& f(0,0,\mathbf{u}) + \mathbf{x} \cdot \nabla f, \\
    g(\mathbf{x},t,\mathbf{u}) &=& g(0,0,\mathbf{u}) + \mathbf{x} \cdot \nabla g+ g_t t.
\end{eqnarray}
Substitute these formulas into the analytical solution,
\begin{eqnarray}
    f_{ugks}(0,t,\mathbf{u}) &=& e^{-t/\tau}(f(0,0,\mathbf{u}) - t\mathbf{u}\cdot \nabla f) + (1-e^{-t/\tau})g(0,0,u) \nonumber\\
    && +(-\tau+(\tau+t)e^{-t/\tau}) \mathbf{u}\cdot \nabla g + (t-\tau+\tau e^{-t/\tau}) g_t. \nonumber
\end{eqnarray}
This is the distribution function at the cell interface. The numerical microscopic flux is,
\begin{eqnarray}
    \mathcal{F}_{ugks} &=& \int_{0}^{\Delta t} u_k f_{ugks}(0,t,\mathbf{u}_k) dt. \label{eq:ugksflux0}
\end{eqnarray}
Then taking moments of above solution, we get the numerical macroscopic flux at cell interface.
\begin{eqnarray}
    \mathcal{F}^W_{ugks} &=& <\mathcal{F}_{ugks}>_k.
\end{eqnarray}

As a standard finite volume method, the quantities inside a cell are updated by considering both the numerical flux and the body force.
Because of the conservation constraint on the collision term, the source terms for conservative variables are zero,
\begin{eqnarray}
  \langle \psi(f-g) \rangle = 0 \quad \mbox{or} \quad \langle \psi(f-g) \rangle_k = 0.
\end{eqnarray}
Therefore, the conservative variables can be updated by only taking account of the numerical macroscopic flux,
\begin{eqnarray}
  W^{n+1} = W^{n} - \nabla\cdot \mathcal{F}^W. \label{eq:macroConserve}
\end{eqnarray}
After obtaining $W^{n+1}$, the equilibrium state $g^{n+1}$ is known through the formula (Eq.(\ref{eq:equilibrium})). The time discretization of the kinetic equation (Eq.(\ref{eq:BGK})) can be written as,

\begin{eqnarray}
\frac{f_{k}^{n+1}-f_{k}^{n}}{\Delta t} + \frac{1}{\Delta t}\nabla\cdot \mathcal{F}_{k} = \frac{g_{k}^{n+1} - f_{k}^{n+1}}{\tau}.
\end{eqnarray}
Then solve the distribution function at $n+1$ step,
\begin{eqnarray}
f_{k}^{n+1} = \frac{\tau}{\tau+\Delta t}(f_{k}^{n}-\nabla\cdot\mathcal{F}_{k}) + \frac{\Delta t}{\tau+\Delta t}g^{n+1}_{k}. \label{eq:lastTimeStepCE}
\end{eqnarray}

As shown above, the convection term $\nabla \mathcal{F}$ is also considered when evaluating the body force. The strong coupling of collision and convection term in the scheme is the main distinguishing feature compared to the operator splitting DOM. The UGKS take the complete equation to evaluate the numerical flux and the body force. This is the reason why the UGKS is an NS AP scheme.

In this study, we use 'DOM' to denote the numerical scheme which couples the collisionless flux (Eq.(\ref{eq:domflux})) and the implicit time discretization (Eq.(\ref{eq:lastTimeStepCE},\ref{eq:domfluxW})) for the body force.
The time discretization (Eq.(\ref{eq:lastTimeStepCE})) is adopted as a common ingredient of all the numerical schemes compared in this paper.

\subsection{The numerical fluxes stem from the equilibrium and non-equilibrium parts}

The numerical fluxes of the unified gas kinetic scheme are composed of the equilibrium and non-equilibrium terms. The competition of all these terms determines the asymptotic behavior of the numerical schemes in different flow regimes. This issue has been discussed by Mieussens \cite{Mieussens2013} for the UGKS of radiative transfer equation.
We will investigate every term in detail and deduce the asymptotic coefficient of each term.
The numerical flux (Eq.(\ref{eq:ugksflux0})) can be further unfolded as follows,
\begin{eqnarray}
\mathcal{F}_{ugks} &=& u_k \{\gamma_0^{ugks}f_k + \gamma_1^{ugks} \mathbf{u}_k\cdot \nabla f_k \nonumber\\
 && +\gamma_2^{ugks}g_k +\gamma_3^{ugks} (\mathbf{u}_k\cdot \nabla g_k + \frac{\partial g_k}{\partial t}) + \gamma_4^{ugks} \frac{\partial g_k}{\partial t}\}, \label{eq:ugksflux}\\
\mathcal{F}^W_{ugks} &=& \gamma_0^{ugks}\langle u \psi f \rangle_k + \gamma_1^{ugks}\langle u \psi \mathbf{u}\cdot \nabla f \rangle_k \nonumber\\
 && +\gamma_2^{ugks}\langle u \psi g \rangle_k + \gamma_3^{ugks}\langle u \psi (\mathbf{u}\cdot \nabla g +g_t) \rangle_k + \gamma_4^{ugks}\langle u \psi g_t \rangle_k.\label{eq:ugksfluxW}
\end{eqnarray}
For the sake of simplicity, we define the coefficients in the UGKS flux as follows,
\begin{eqnarray}
\gamma^{ugks}_0 &=& \tau(1-e^{-\beta}), \nonumber\\
\gamma^{ugks}_1 &=& - \tau(-\Delta t e^{-\beta}+\tau-\tau e^{-\beta}), \nonumber\\
\gamma^{ugks}_2 &=& \Delta t - \tau(1-e^{-\beta}), \\
\gamma^{ugks}_3 &=& \tau(-\Delta t e^{-\beta} - 2\tau e^{-\beta} +2\tau-\Delta t), \nonumber\\
\gamma^{ugks}_4 &=& (\frac{1}{2}\Delta t^2 + \tau((\Delta t +\tau) e^{-\beta} -\tau)), \nonumber
\end{eqnarray}
where $\beta$ is defined as the ratio of the time step $\Delta t$ to the relaxation time $\tau$, namely, $\beta = \Delta t/\tau$.
The first two terms on the right hand side of the Eq.(\ref{eq:ugksflux}) are the non-equilibrium parts which are deduced from the non-equilibrium initial condition at the beginning of the time step.
The last three terms on the right hand side stemming from the collision term represent the Navier-Stokes flux.
As shown above, the non-equilibrium part $f_0$ does not vanish directly when $\frac{\Delta t}{\tau}\rightarrow +\infty$. A small term ($O(\tau)$) still influences the numerical fluxes. Xu provided a profound perspective of the asymptotic behavior of the numerical flux \cite{Xu2001}. He showed that proper initial condition (Chapman-Enskog expansion) of each time step should be assumed to deduce correct numerical flux in the continuum flow regimes. Following this idea, we consider a specific expression of the non-equilibrium part. For the sake of the implicit discretization (Eq.(\ref{eq:lastTimeStepCE})) of the collision term, the following assumption seems natural and rational. The initial condition $f$ deviates from the equilibrium by $O(\tau)$, namely,
\begin{eqnarray}
  f = g+O(\tau). \label{eq:assumptionO}
\end{eqnarray}
After some derivations, we can get more precise estimation for the initial condition \cite{chen2015comparative}, that is,
\begin{eqnarray}
    \begin{array}{ccl}
    f(0,0,\mathbf{u}) &=& g(0,0,\mathbf{u}) - \tau(\mathbf{u}\cdot \nabla g + g_t) + O(\tau\Delta t), \\
    \end{array} \label{eq:assumptionCE}
\end{eqnarray}
where we choose the approximate Chapman-Enskog expansion (Eq.(\ref{eq:assumptionCE})) as the initial condition for the UGKS.
Then substituting the estimation (Eq.(\ref{eq:assumptionCE})) into the Eq.(\ref{eq:ugksflux}) and Eq.(\ref{eq:ugksfluxW}), the numerical flux becomes,
\begin{eqnarray}
  \mathcal{F}_{ugks} &=& u_k(\Delta t g_k - \Delta t \tau(\mathbf{u}_k\cdot \nabla g_k + \frac{\partial g_k}{\partial t}) + \frac{\Delta t^2}{2}\frac{\partial g_k}{\partial t}) +O(\tau^2), \nonumber \\
  \mathcal{F}^W_{ugks} &=& \langle u\psi(\Delta t g - \Delta t \tau(\mathbf{u}\cdot \nabla g + \frac{\partial g}{\partial t}) + \frac{\Delta t^2}{2}\frac{\partial g}{\partial t}) \rangle_k + O(\tau^2).\label{eq:ugksCE}
\end{eqnarray}
The Chapman-Enskog expansion for the Navier-Stokes equation is exactly recovered. Please note that, $\beta$ is not required to approach zero as we derive the Chapman-Enskog expansion. As we know, the numerical scheme must converge as the time step goes to zero. In this sense, the asymptotic behavior when $\tau\rightarrow 0,\ \Delta t\rightarrow 0$, and $\beta$ is finite, is more important to the numerical scheme.

Under the more precise assumption (Eq.(\ref{eq:assumptionCE})), the estimation of the numerical fluxes in the DOM is written as,
\begin{eqnarray}
  \mathcal{F}_{dom} &=& u_k(\Delta t g_k - \Delta t(\tau+\frac{\Delta t}{2})(\mathbf{u}_k\cdot \nabla g_k + \frac{\partial g_k}{\partial t}) + \frac{\Delta t^2}{2}\frac{\partial g_k}{\partial t}) +O(\tau^2), \nonumber\\
  \mathcal{F}^W_{dom} &=& \langle u \psi(\Delta t g - \Delta t(\tau+\frac{\Delta t}{2})(\mathbf{u}\cdot \nabla g + \frac{\partial g}{\partial t}) + \frac{\Delta t^2}{2}\frac{\partial g}{\partial t})\rangle_k + O(\tau^2).\label{eq:domCE}
\end{eqnarray}
The equivalent viscosity in Eq.(\ref{eq:domCE}) is enlarged by the free streaming. We use $\alpha_{dom}$ to denote the enlarging factor, which is
$$\alpha_{dom} = (\tau+\Delta t/2)/\tau = 1+\frac{1}{2}\beta.$$
It is close to KFVS-NS for a discontinuous flow. When $\beta$ varies from 0 to $\infty$, $\alpha_{dom}$ diverges . The enlarged viscosity is an analogue to the numerical viscosity in the lattice Boltzmann method \cite{chen1992recovery} before the remedy of the viscosity.

\section{Simplification of the unified gas kinetic scheme}
In the UGKS fluxes (Eq.(\ref{eq:ugksflux},\ref{eq:ugksfluxW})), the last three terms which stem from the collision term, are also discretized in velocity space. Therefore, it takes huge computational resources compared to the traditional
Navier-Stokes solvers. In fact, more than half portion of computation resource is taken to evaluate the equilibrium part. Actually, the quadratures£¬ $\langle g \rangle_k, \langle \nabla g\rangle_k$, and $\langle g_t \rangle_k$ are only approximation of $\langle g \rangle, \langle \nabla g\rangle$, and $\langle g_t \rangle$. The quadrature of $g$ and its derivatives can be calculated analytically, for instance, $\langle g\rangle = (\rho, \rho U, \rho E)^T$. If the quadratures of the equilibrium state $g$ and its derivatives are handled in traditional way in terms of analytical macroscopic flux \cite{Xu2001}, the unified scheme will be much more efficient. Therefore, we propose the first simplification (S1), that is, using traditional DOM to calculate the flux for distribution function and using the macroscopic gas kinetic scheme \cite{Xu2001} to evaluate the last three terms in Eq.(\ref{eq:ugksfluxW}),
\begin{eqnarray}
  \mathcal{F}_{s1} &=& u_k (\Delta t f_{k} - \frac{1}{2}\Delta t^2 \mathbf{u}_k\cdot \nabla f_{k}), \nonumber\\
  \mathcal{F}_{s1}^W &=& \gamma_0^{s1}\langle u \psi f \rangle_k + \gamma_1^{s1}\langle u \psi \mathbf{u}\cdot \nabla f \rangle_k \nonumber\\
 && +\gamma_2^{s1}\langle u \psi g \rangle + \gamma_3^{s1}\langle u \psi (\mathbf{u}\cdot \nabla g +g_t) \rangle + \gamma_4^{s1}\langle u \psi g_t \rangle. \label{eq:s1flux} \\
  \gamma_i^{s1} &=& \gamma_i^{ugks}, \quad i = 0,1,2,3,4. \nonumber
\end{eqnarray}
Compared to the numerical macroscopic flux of the UGKS (Eq.(\ref{eq:ugksfluxW})), the equilibrium part is solved analytically (note the different symbols $\langle \cdot \rangle$ and $\langle \cdot \rangle_k$), and the numerical microscopic flux (Eq.(\ref{eq:ugksflux})) is replaced by the traditional DOM (Eq.(\ref{eq:domflux})). With the assumption (Eq.(\ref{eq:assumptionCE})), if the difference between $\langle\cdot\rangle$ and $\langle\cdot\rangle_k$ is ignored, the numerical microscopic flux becomes,
\begin{eqnarray}
  \mathcal{F}_{s1} &=& u_k (\Delta t g_{k} - (\tau+\frac{1}{2}\Delta t)\Delta t(\mathbf{u}_k\cdot \nabla g_k + \frac{\partial g_k}{\partial t}) + \frac{1}{2}\Delta t^2 \frac{\partial g_k}{\partial t})+ O(\tau^2), \nonumber\\
  \mathcal{F}^W_{s1} &=& \langle u\psi(\Delta t g - \Delta t \tau(\mathbf{u}\cdot \nabla g + \frac{\partial g}{\partial t}) + \frac{\Delta t^2}{2}\frac{\partial g}{\partial t}) \rangle + O(\tau^2).
\end{eqnarray}
If the quadrature $(\langle\cdot\rangle_k)$ is accurate, the numerical macroscopic flux of the S1 scheme is identical to the macroscopic flux of the UGKS. Only the flux for the distribution function is different.
We will present some numerical comparisons to demonstrate that the inaccurate microscopic numerical flux has very little influence to the NS AP property of the numerical scheme.
This simplification only reduces the computational cost, but the formula and the coding are still complicated. Hence, we propose a second simplified method (S2), which is barely a combination of the DOM and the gas kinetic scheme for the Navier-Stokes equations. The numerical fluxes are given as follows.
\begin{eqnarray}
  \mathcal{F}_{s2} &=& u_k (\Delta t f_{k} - \frac{1}{2}\Delta t^2 \mathbf{u}_k\cdot \nabla f_{k}), \nonumber \\
  \mathcal{F}_{s2}^W &=& e^{-\beta} \langle u \psi (\Delta t f - \frac{1}{2}\Delta t^2 \mathbf{u}\cdot \nabla f) \rangle_k \label{eq:s2flux}\nonumber\\
  && + (1-e^{-\beta}) \langle u\psi(\Delta t g_0 - \Delta t\tau(\mathbf{u}\cdot \nabla g + \frac{\partial g}{\partial t}) + \frac{1}{2}\Delta t^2 \frac{\partial g}{\partial t}) \rangle \label{eq:s2flux}
\end{eqnarray}

\begin{table}
\begin{tabular}{|c |c|c|c|}
  \hline
    & formula & $\beta \rightarrow \infty$ & $\beta \rightarrow 0$ \\
  \hline
  \hline
  $\gamma_0^{ugks}$ & $\tau(1-e^{-\beta})$ & $\tau + O(e^{-\beta})$ & $\Delta t -\frac{\Delta t^2}{2\tau} + O(\beta^2)$ \\
  $\gamma_0^{s2}$ & $\Delta t e^{-\beta}$ & $O(e^{-\beta})$ & $\Delta t -\frac{\Delta t^2}{\tau} + O(\beta^2)$ \\
  \hline
  $\gamma_1^{ugks}$ & $- \tau(-\Delta t e^{-\beta}+\tau-\tau e^{-\beta})$ & $-\tau^2 + O(e^{-\beta})$ & $-\frac{\Delta t^2}{2} + \frac{\Delta t^3}{3\tau} + O(\beta^2)$ \\
  $\gamma_1^{s2}$ & $-\frac{1}{2}\Delta t^2 e^{-\beta}$ & $O(e^{-\beta})$ & $-\frac{\Delta t^2}{2} + \frac{\Delta t^3}{2\tau} + O(\beta^2)$ \\
  \hline
  $\gamma_2^{ugks}$ & $\Delta t - \tau(1-e^{-\beta})$ & $\Delta t - \tau + O(e^{-\beta})$ & $\frac{\Delta t^2}{2\tau} + O(\beta^2)$ \\
  $\gamma_2^{s2}$ & $(1-e^{-\beta})\Delta t$ & $\Delta t + O(e^{-\beta})$ & $\frac{\Delta t^2}{\tau}+O(\beta^2)$ \\
  \hline
  $\gamma_3^{ugks}$ & $\tau(2\tau-\Delta t - (\Delta t + 2\tau) e^{-\beta} )$ & $ 2\tau^2 -\tau\Delta t + O(e^{-\beta})$ & $-\frac{\Delta t^3}{6\tau} + O(\beta^2)$ \\
  $\gamma_3^{s2}$ & $-\tau\Delta t(1-e^{-\beta})$ & $-\tau\Delta t + O(e^{-\beta})$ & $\Delta t^2-\frac{\Delta t^3}{2\tau} + O(\beta^2)$ \\
  \hline
  $\gamma_4^{ugks}$ & $\frac{1}{2}\Delta t^2 + \tau((\Delta t +\tau) e^{-\beta} -\tau)$ & $\frac{1}{2}\Delta t^2 - \tau^2 + O(e^{-\beta})$ & $\frac{\Delta t^3}{3\tau} + O(\beta^2)$ \\
  $\gamma_4^{s2}$ & $\frac{1}{2}\Delta t^2(1-e^{-\beta})$ & $\frac{1}{2}\Delta t^2+O(e^{-\beta})$ & $\frac{\Delta t^3}{2\tau} + O(\beta^2)$ \\
  \hline
\end{tabular}
\caption{The coefficients of the numerical flux in UGKS and S2 scheme, where $\beta = \Delta t/\tau$.} \label{tab:coe}
\end{table}

This method is very simple. We can easily combine two existing flux solvers to construct a unified scheme for gas kinetic equation.
Assume that the initial condition at the beginning of the time step satisfies the near equilibrium assumption, namely, Eq.(\ref{eq:assumptionCE}) is applied. The numerical flux of the second simplified method becomes,
\begin{eqnarray}
  \mathcal{F}_{s2} &=& u_k (\Delta t g_{k} - (\tau+\frac{\Delta t}{2})\Delta t(\mathbf{u}_k\cdot \nabla g_k + \frac{\partial g_k}{\partial t}) + \frac{1}{2}\Delta t^2 \frac{\partial g_k}{\partial t})+ O(\tau^2), \nonumber\\
  \mathcal{F}^W_{s2} &=& \langle u\psi(\Delta t g - \Delta t (\tau-\frac{\Delta t}{2} e^{-\beta})(\mathbf{u}\cdot \nabla g + \frac{\partial g}{\partial t}) + \frac{1}{2}\Delta t^2\frac{\partial g}{\partial t}) \rangle+ O(\tau^2). \nonumber
\end{eqnarray}
Obviously, in the continuum flow regime, this simplification is accurate enough to lead to the Navier-Stokes numerical flux.
The viscosity of the S2 scheme is enlarged by a factor,
\begin{eqnarray}
  \alpha_{s2} = (\tau-\Delta t e^{-\beta}/2)/\tau = 1-\frac{1}{2}\beta e^{-\beta}.
\end{eqnarray}
It only varies inside the interval $[1-e^{-1}/2,1]$. The minimum is attained when $\beta = 1$.
The table \ref{tab:coe} compares the coefficients of the UGKS and the second simplified scheme. As shown in the second column, the formulas are apparently different when $\beta$ has a finite value. When $\beta$ goes to infinity and $\tau$ goes to zero, i.e., in the continuum flow regime, the coefficients are identical up to $O(\tau)$. The S2 scheme approaches the equilibrium state a little faster than the UGKS because the coefficients of the non-equilibrium part, $\gamma_0^{s2}$ and $\gamma_1^{s2}$, approach to zero more rapidly. Consider the free molecular flow limit, namely, $\beta$ goes to zero and $\tau$ goes to infinity. The coefficients of the UGKS and the S2 scheme are identical up to $O(\beta)$, except $\gamma_3^{s2}$. It deviates from $\gamma_3^{ugks}$ in free molecular flow regime largely.
This means the simple combination cannot recover the free molecular flow regime. We will also find large discrepancy generated from $\gamma_3^{s2}$ in the numerical comparison section.

\begin{figure}
\centering
    %\parbox[b]{0.6\textwidth}{
%\centering
    \includegraphics[totalheight=5cm]{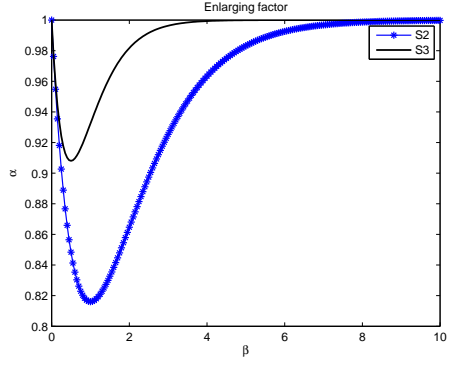}
    %}
    \caption{The enlarging factor of the viscosity for the simplified scheme (S2, S3).}
    \label{fig:enlargingFactor}
\end{figure}

Therefore, we propose a third simplified method (S3) which modified the coefficient in front of the Navier-Stokes viscous term.
The basic idea is to construct a coefficient $\gamma_3^{s3}$ which can preserve the asymptotic limit of $\gamma^{ugks}_3$.
The third simplified method is,
\begin{eqnarray}
  \mathcal{F}_{s3} &=& u_k (\Delta t f_{k} - \frac{1}{2}\Delta t^2 \mathbf{u}_k\cdot \nabla f_{k}) \nonumber\\
  \mathcal{F}_{s3}^W &=& e^{-\beta}\langle u \psi (\Delta t f - \frac{1}{2}\Delta t^2 \mathbf{u}\cdot \nabla f)\rangle_k \nonumber\\
  && + (1-e^{-\beta}) \langle u\psi(\Delta t g_0 - \Delta t r_{\tau} \tau  (\mathbf{u}\cdot \nabla g + \frac{\partial g}{\partial t}) + \frac{1}{2}\Delta t^2 \frac{\partial g}{\partial t}) \rangle \label{eq:s3flux}\\
  r_{\tau} &=& \frac{1-e^{-\beta}-\frac{\Delta t}{2\tau}(1+e^{-\beta})e^{-\beta}}{1-e^{-\beta}}. \label{eq:rTau}
\end{eqnarray}
The coefficient $\gamma_3^{s3}$ becomes,
\begin{eqnarray}
  \gamma_3^{s3} &=& -\tau\Delta t(1-e^{-\beta}-\frac{\Delta t}{2\tau}(1+e^{-\beta})e^{-\beta}), \nonumber\\
  \gamma_3^{s3} &=& -\tau\Delta t + O(e^{-\beta}),\quad \mbox{when}\ \frac{\Delta t}{\tau}\rightarrow \infty \nonumber\\
  \gamma_3^{s3} &=& -\frac{\Delta t^3}{\tau}+O(\beta^2),\quad \mbox{when}\ \frac{\Delta t}{\tau}\rightarrow 0 \nonumber
\end{eqnarray}
The only difference from the second simplified method is that, the coefficient in front of the viscous term is multiplied by a factor $r_{\tau}$.
Therefore, the third method can also be taken as a simple combination between the DOM and the Navier-Stokes solver.
As we can see, the coefficient $\gamma_3^{s3}$ has the same limit in free molecular flow regime up to $O(\beta)$.
Then considering the continuum flow regime, with the assumption Eq.(\ref{eq:assumptionCE}), the third simplified method becomes,
\begin{eqnarray}
  \mathcal{F}_{s3} &=& u_k (\Delta t g_{k} - (\tau+\frac{1}{2}\Delta t)\Delta t(\mathbf{u}_k\cdot \nabla g_k + \frac{\partial g_k}{\partial t}) + \frac{1}{2}\Delta t^2 \frac{\partial g_k}{\partial t})+ O(\tau^2), \nonumber\\
  \mathcal{F}^W_{s3} &=& \langle u\psi(\Delta t g - \Delta t \tau(\mathbf{u}\cdot \nabla g + \frac{\partial g}{\partial t}) + \frac{\Delta t^2}{2}\frac{\partial g}{\partial t}) \rangle+ O(\tau^2) \nonumber\\
  &&- \langle u\psi(\Delta t (\frac{1}{2}\Delta t e^{-\beta}+\tau(1-e^{\beta})(r_{\tau}-1))(\mathbf{u}\cdot \nabla g + \frac{\partial g}{\partial t}))\rangle \\
  &=& \langle u(\Delta t g - \Delta t (\tau-\frac{\Delta t}{2} e^{-2\beta})(\mathbf{u}\cdot \nabla g + \frac{\partial g}{\partial t}) + \frac{\Delta t^2}{2}\frac{\partial g}{\partial t}) \rangle+ O(\tau^2).  \nonumber
\end{eqnarray}
For the third simplified method, the equivalent viscosity is enlarged by,
\begin{eqnarray}
  \alpha_{s3} = (\tau-\Delta t e^{-2\beta}/2)/\tau =1-\frac{1}{2}\beta e^{-2\beta}.
\end{eqnarray}
It only varies inside the interval $[1-e^{-1}/4,1]$. The minimum is attained when $\beta = 1/2$. Figure \ref{fig:enlargingFactor} shows the enlarging factor $\alpha$ versus $\beta$. The numerical flux of the UGKS is based on the analytical solution. Therefore, its viscosity is unchanged in the second order temporal discretization (Eq.(\ref{eq:ugksCE})), i.e., $\alpha_{ugks}=1$. The simplified schemes somehow modify the viscous coefficient. As shown in the figure, the S3 scheme is more accurate than the S2 scheme in terms of the viscosity coefficient.

We analyze the behavior of these simplified numerical schemes. The S1 scheme replaces the quadrature related to the equilibrium state by the analytical solution. Although it has correct asymptotic limits and less computational cost, the scheme is still complicated in terms of coding. The S2 scheme is a simple combination of Navier-Stokes solver and traditional DOM. It cannot reproduce the free molecular flow regime. The S3 scheme has correct asymptotic limits in free molecular flow regime, and also in the continuum flow regime. For the transition flow regime, the coefficients are apparently different from the analytical solution. We will use the numerical experiment to investigate the performance of different simplifications.

\section{Numerical discretization}
The previous section introduced the numerical flux expression in terms of time. Several simplified numerical fluxes are constructed based on the unified gas kinetic scheme. In this section the spatial discretization and the boundary condition are provided.
\subsection{Spatial discretization}
The value and its spatial derivative of a certain quantity are needed in the expressions of the numerical flux (for example Eq. (\ref{eq:ugksflux})). For the velocity distribution function, we adopt the third order WENO to interpolate its value at the cell interface $(i+1/2)$, where $i$ denotes the index along the interpolation direction. The formula is given below,
\begin{eqnarray}
f_{l} = \frac{w_{-1} f^{-1} + w_{0}f^{0}}{w_{-1}+w_0}, \quad f_{r} = \frac{w_0 f^0 + w_{1}f^{1}}{w_0+w_{1}}, \nonumber
\end{eqnarray}
where the subscript '$l$' and '$r$' represent left side and right side respectively, and $w$ denotes the weight. Their formulas are written as follows,
\begin{eqnarray}
w_{-1} = \frac{1}{4(s_{i-1}^2+\varepsilon)},\quad w_0 = \frac{3}{4(s_i^2+\varepsilon)}, \quad w_{1} = \frac{1}{4(s_{i+1}^2+\varepsilon)}, \nonumber
\end{eqnarray}
where $\varepsilon = 1\times 10^{-6}$ is used to prevent zero denominator, and $s_i = f_{i+1}-f_{i}$,
\begin{eqnarray}
f^{-1} = \frac{3}{2}f_{i}-\frac{1}{2}f_{i-1},\quad f^0 = \frac{1}{2}f_{i+1}+\frac{1}{2}f_{i}, \quad f^{1} = \frac{3}{2}f_{i+1}-\frac{1}{2}f_{i+2} \nonumber
\end{eqnarray}

For high speed flow, the 3rd order WENO is also employed to calculate the macroscopic variables at the cell interface, owing to the discontinuous shock wave in the flow field.
For low speed flow, the macroscopic conservative variables are interpolated by the central difference method, that is,
\begin{eqnarray}
W_{i+1/2} = \frac{1}{2}(W_{i}+W_{i+1}).
\end{eqnarray}

The derivatives of the microscopic and macroscopic variables are evaluated by a second order central difference method.

\subsection{Boundary condition}
Boundary condition is another crucial ingredient for AP schemes. At first, we recall the diffusion boundary condition for the traditional DOM in free molecular flow regime.
The distribution function of the reflecting particles is subjected to the Maxwell distribution. Since no penetration occurs during the collision with the wall, the mass flux of the particle can be written as follows,
\begin{eqnarray}
  \int_0^{\Delta t}\int_{0}^{+\infty} u f^{in} d\mathbf{u}dt + \rho_{dom}\int_0^{\Delta t}\int_{-\infty}^{0} u g^* d\mathbf{u}dt = 0, \label{eq:nopenetration}
\end{eqnarray}
where $f^{in}$ represents the incident molecular distribution function which is interpolated from the interior of the flow field, $\rho_{dom}$ is the density of the reflecting molecular stream. The reflecting molecular distribution function is assumed to be the Maxwell equilibrium on the wall, which reads,
\begin{eqnarray}
  g^* = \left(\frac{2RT^*}{\pi}\right)^{3/2} e^{-\frac{1}{2RT^*}\mathbf{u}^2},    \label{eq:wallEquilibrium}
\end{eqnarray}
where $T^*$ denote the temperature of the boundary.
According to Eq.(\ref{eq:nopenetration}), the density of the reflecting distribution is determined, that is,
\begin{eqnarray}
  \rho_{dom} = -\frac{\sum\limits_{u_k>0} \omega_k u_k f_k}{\sum\limits_{u_k\leq 0}\omega_k u_k g_k}.
\end{eqnarray}
The velocity distribution function at the wall for the microscopic variables is,
\begin{eqnarray}
  f_{dom} = \left\{
    \begin{array}{c}
      f^{in}, \quad u > 0, \\
      \rho_{dom} g^*, \quad u \leq 0.
    \end{array}\right.
\end{eqnarray}
The numerical fluxes are written as follows,
\begin{eqnarray}
\left\{\begin{array}{ccl}
  \displaystyle\mathcal{F^*}_{dom} = u_k f_{dom,k}, \\
  \displaystyle\mathcal{F^*}^W_{dom} = \langle u \psi f_{dom} \rangle_k.
\end{array}\right.
\end{eqnarray}
The diffusion boundary condition is valid in free molecular flow regime, but cannot automatically recover the no slip boundary condition in the continuum flow regime.
The boundary condition for the simplified method (S2, S3) should be designed carefully to preserve the asymptotic limits. Fortunately, this task is very easy to fulfill, since the simplified scheme is a simple combination of existing schemes. Here we just combine the boundary flux of the diffusion boundary condition and the boundary flux of the gas kinetic scheme for the Navier-Stokes equations to develop a boundary condition for the simplified scheme.

We modify the non-equilibrium bounce back boundary condition \cite{guo2002extrapolation} to implement the isothermal boundary condition for gas kinetic scheme. %treated as the traditional gas kinetic scheme for NS equations . %
We adopt the extrapolation from the interior, then construct the NS distribution at the cell interface as the incident distribution function.
\begin{eqnarray}
  f^{in}_{gks} = g^{in}-r_{\tau}\tau(\mathbf{u}\cdot \nabla g^{in} + g^{in}_t) + g^{in}_t t,  \quad\mbox{for}\  u > 0,
\end{eqnarray}
where $r_{\tau}$ is defined in Eq.(\ref{eq:rTau}).
The reflecting distribution function is constructed as follows,
\begin{eqnarray}
  f^{out}_{gks}(u) = 2\rho_{gks} g^*(u) - f^{in}_{gks}(-u), \quad\mbox{for}\ u \leq 0.
\end{eqnarray}
Then the complete velocity distribution function in the gas kinetic scheme is
\begin{eqnarray}
  f_{gks} = \left\{
    \begin{array}{c}
      f_{gks}^{in}, \quad u > 0, \\
      f_{gks}^{out}, \quad u \leq 0.
    \end{array}\right.
\end{eqnarray}
The no penetration condition is also employed to determine the density at the wall boundary.
\begin{eqnarray}
  \rho_{gks} = \sqrt{\frac{2\pi}{RT^*}}\int_0^{+\infty} u f^{in}_{gks} d\mathbf{u}.
\end{eqnarray}

The numerical flux for the conservative variables and for the distribution function are given respectively.
\begin{eqnarray}
  \left\{\begin{array}{ccl}
  \mathcal{F}_k &=& \mathbf{u}_k f_{dom,k}, \\
  \mathcal{F}^W &=& e^{-\beta} \sum\limits_k u_k \psi_k f_{dom} + (1-e^{-\beta}) <u \psi f_{gks}>. \label{eq:simplifiedBC}
  \end{array}\right.
\end{eqnarray}

We have tested another choice of $r_{\tau}$, say, $r_{\tau} = 1$ for the second simplified method (S2). When applying this boundary condition, in the free molecular flow regime, there were large oscillation near the boundary, since the coefficient $\gamma_3^{s2}$ is inconsistent with the analytical solution ($\gamma_3^{ugks}$). Therefore, only the simplified boundary condition (Eq.(\ref{eq:simplifiedBC})) is adopted for all the numerical simulations in the next section.

\begin{figure}
\begin{center}
    \mbox{Kn=1.0}
    \parbox[b]{0.95\textwidth}{
    \centering
    \includegraphics[totalheight=6.5cm, bb = 92 42 672 545, clip=true]{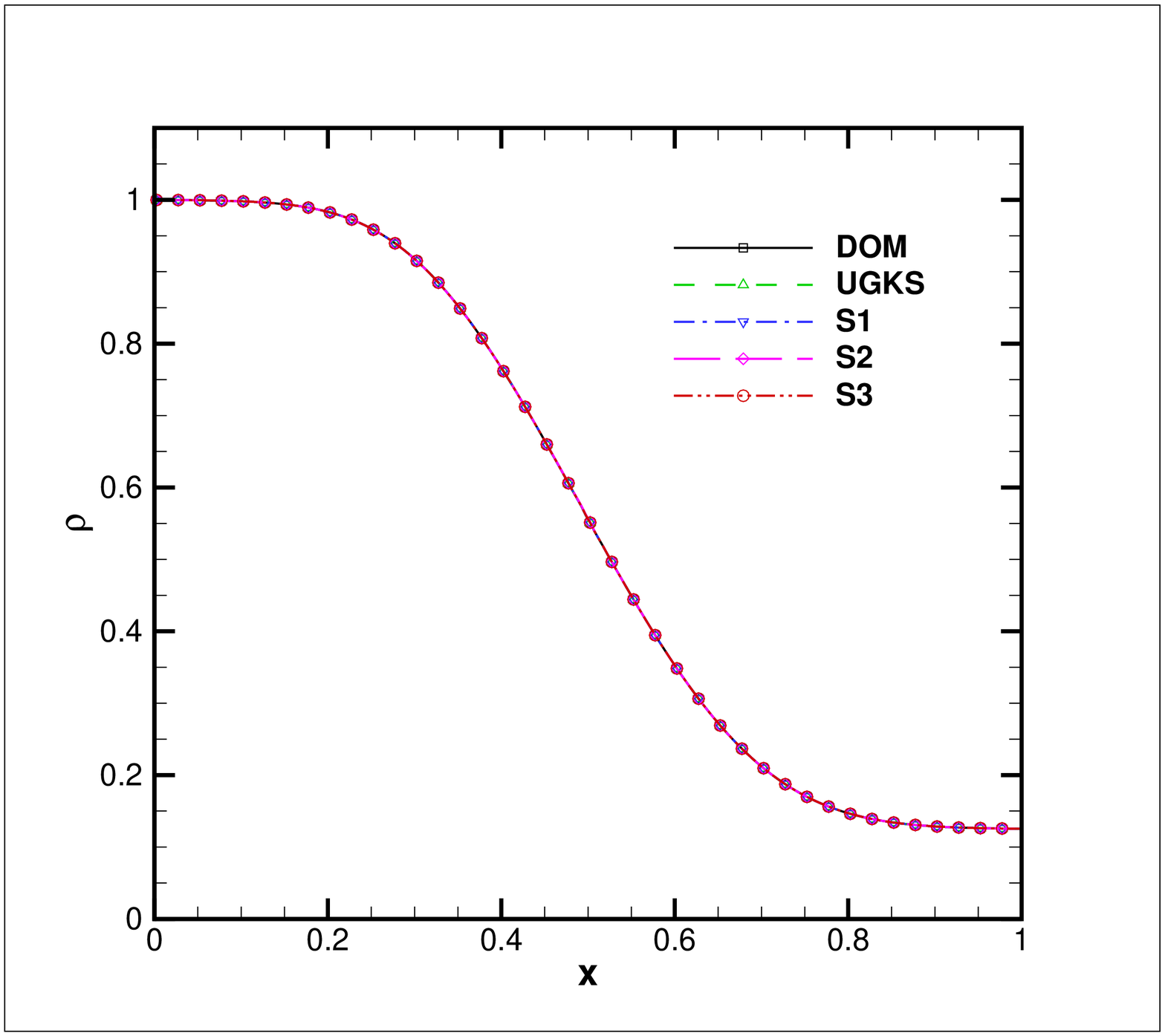}\hfill
    \includegraphics[totalheight=6.5cm, bb = 92 42 672 545, clip=true]{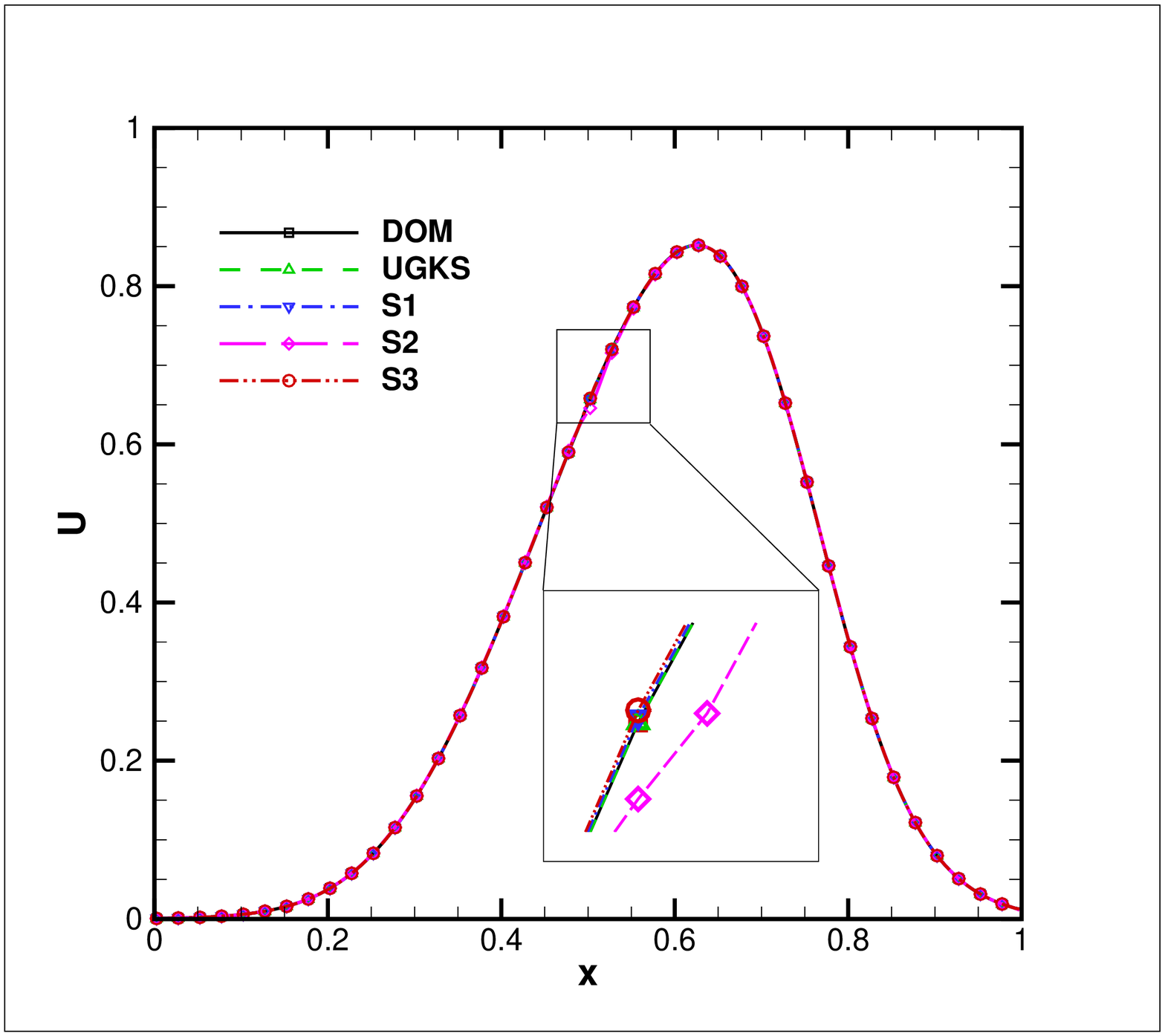}
    }
    \mbox{Kn=0.01}
    \parbox[b]{0.95\textwidth}{
    \centering
    \includegraphics[totalheight=6.5cm, bb = 92 42 672 545, clip=true]{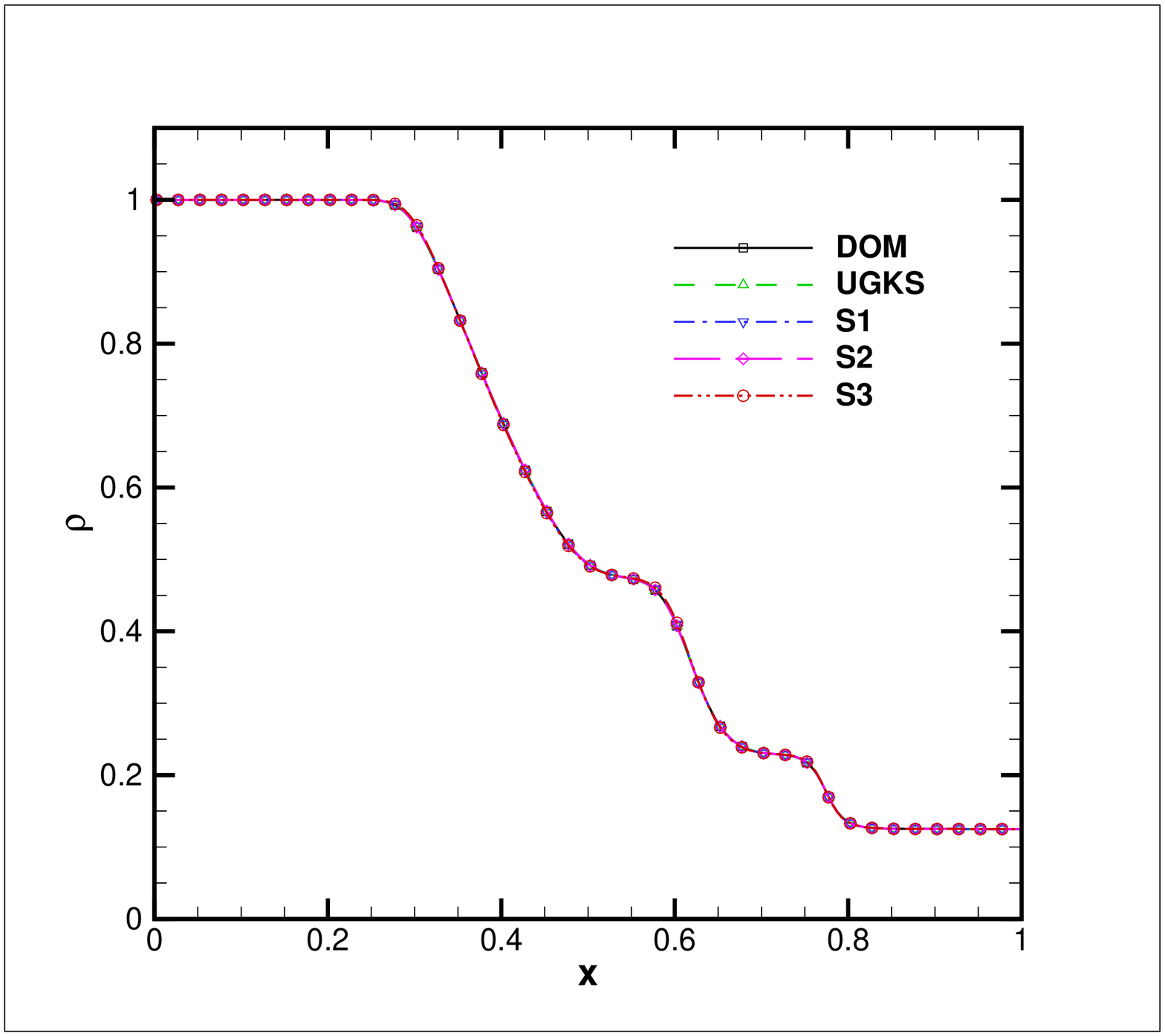}\hfill
    \includegraphics[totalheight=6.5cm, bb = 92 42 672 545, clip=true]{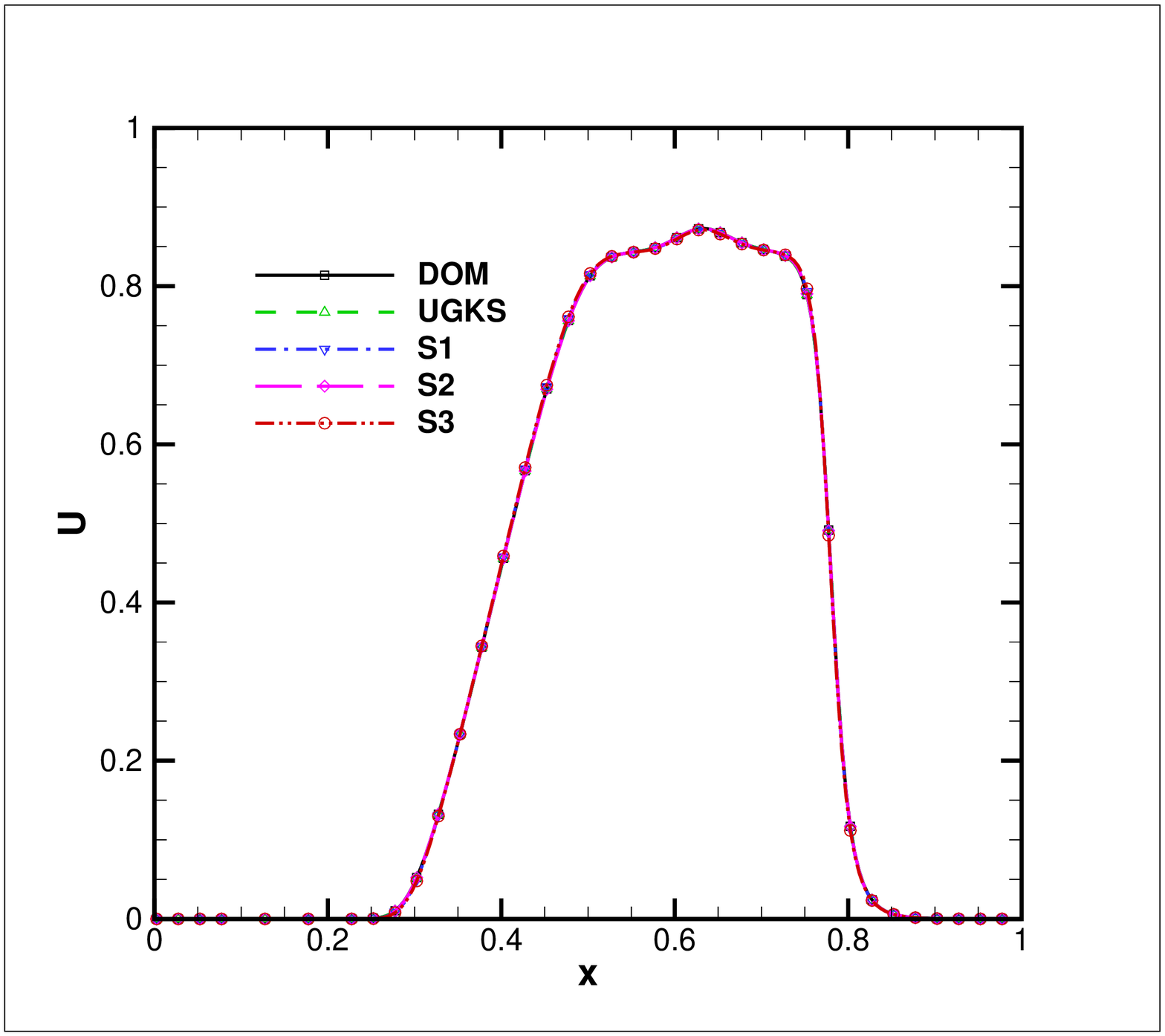}
    }
    \mbox{Kn=0.0001}
    \parbox[b]{0.95\textwidth}{
    \centering
    \includegraphics[totalheight=6.5cm, bb = 92 42 672 545, clip=true]{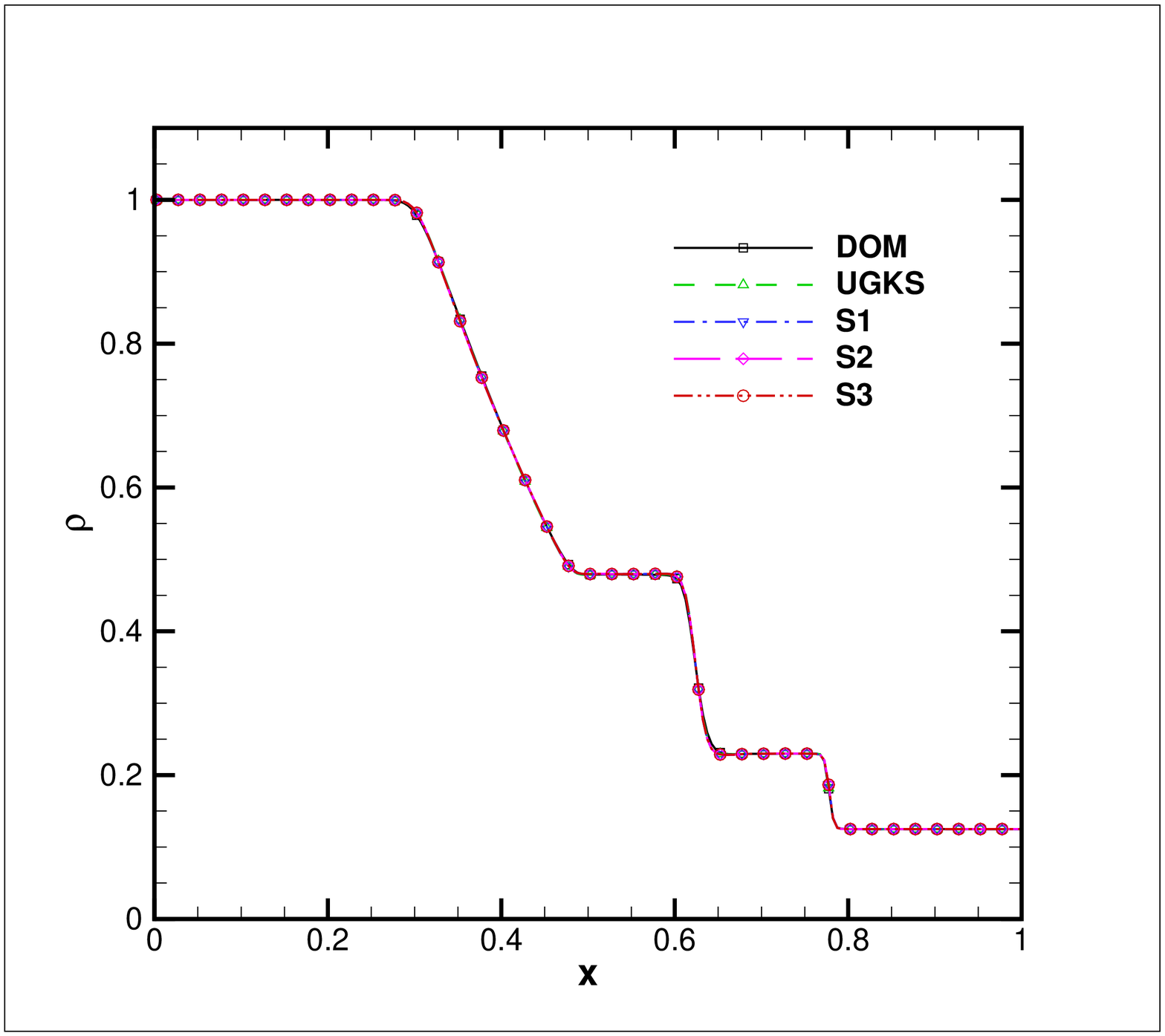}\hfill
    \includegraphics[totalheight=6.5cm, bb = 92 42 672 545, clip=true]{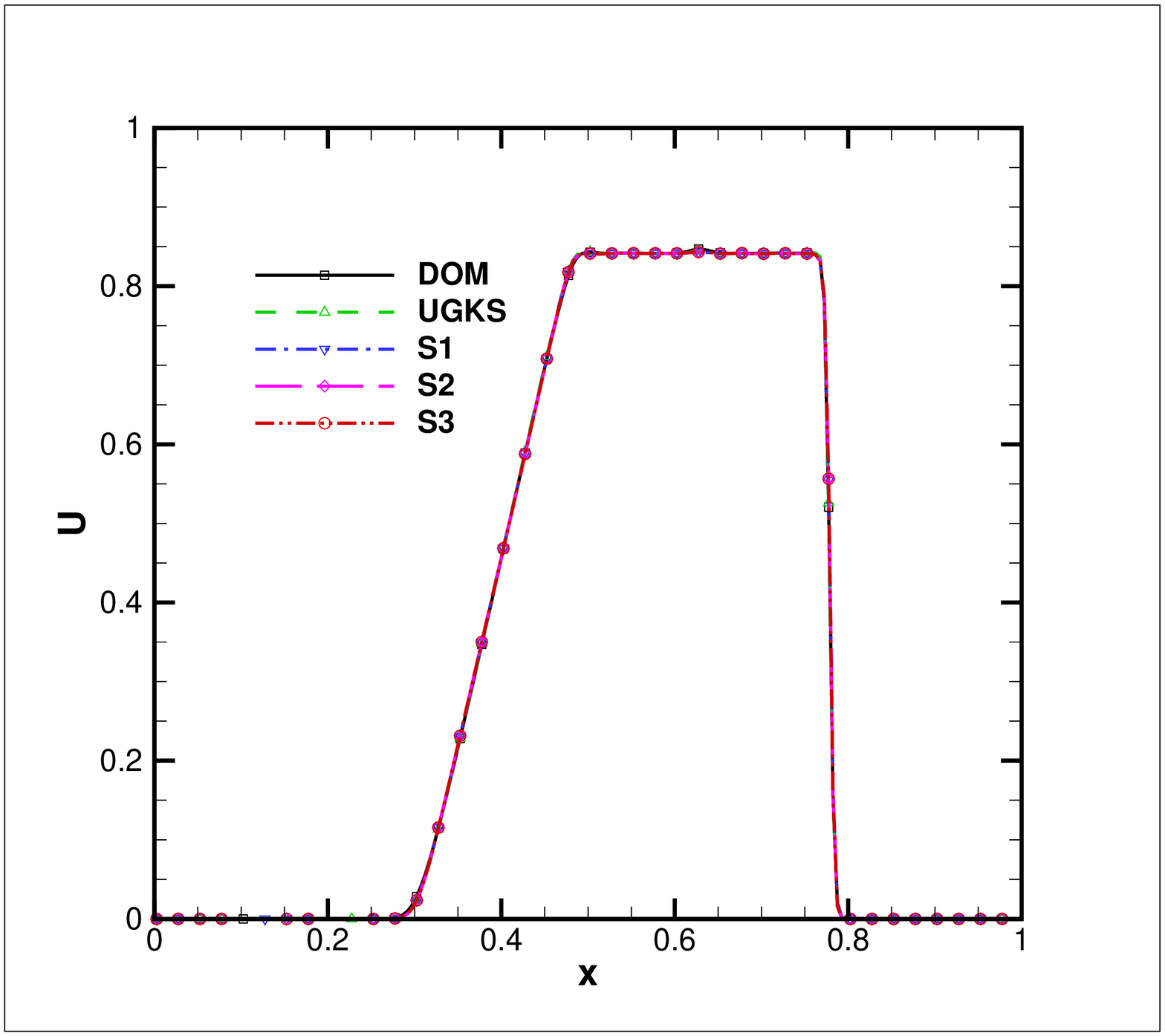}
    }
    \caption{The density and velocity profile of the shock tube problem at different Knudsen number.}
    \label{fig:sod1}
\end{center}
\end{figure}

\begin{figure}
\centering
    \mbox{Kn=10}
    \parbox[b]{0.95\textwidth}{
    \centering
    \includegraphics[totalheight=6.5cm, bb = 92 42 672 545, clip=true]{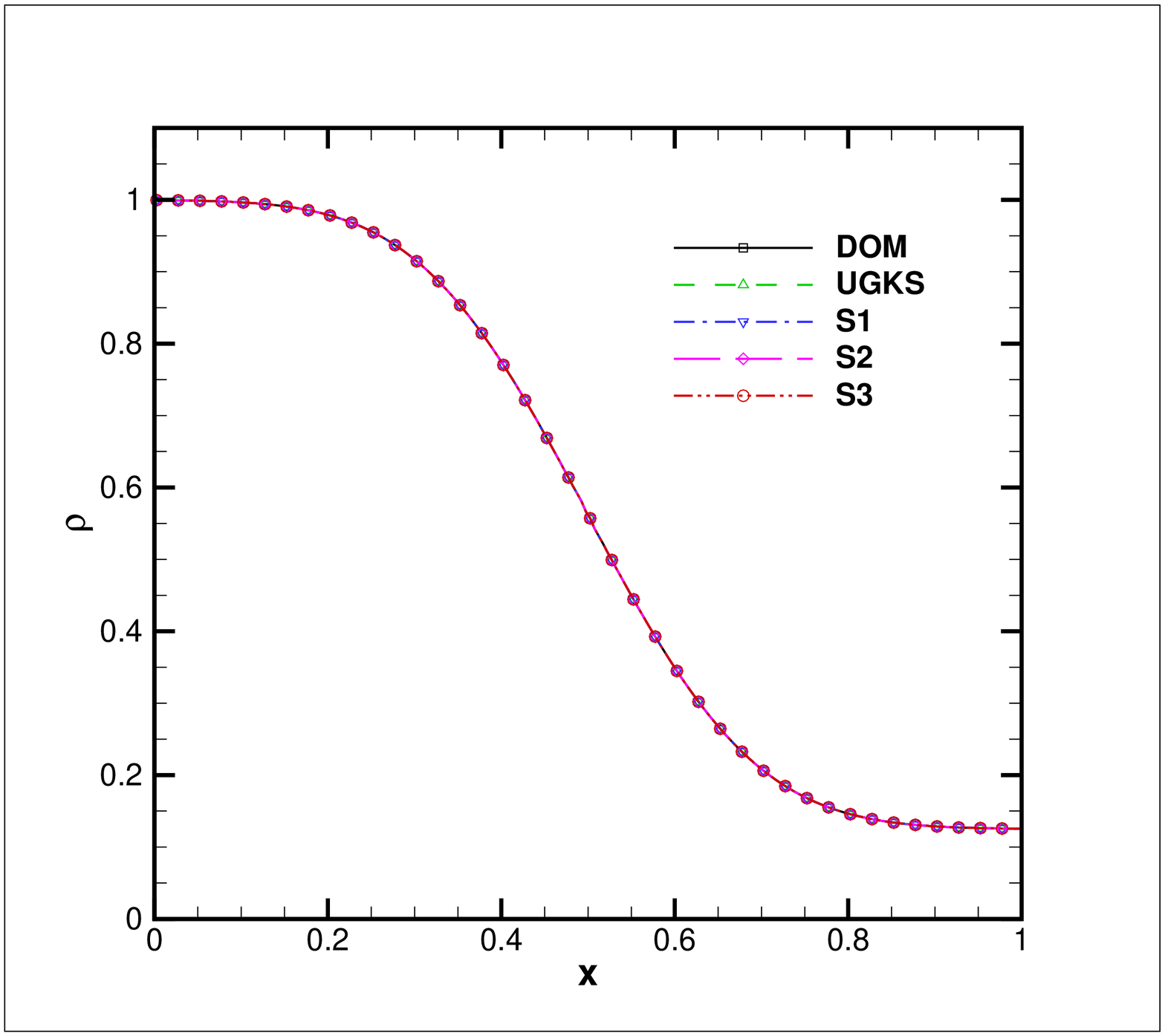}\hfill
    \includegraphics[totalheight=6.5cm, bb = 92 42 672 545, clip=true]{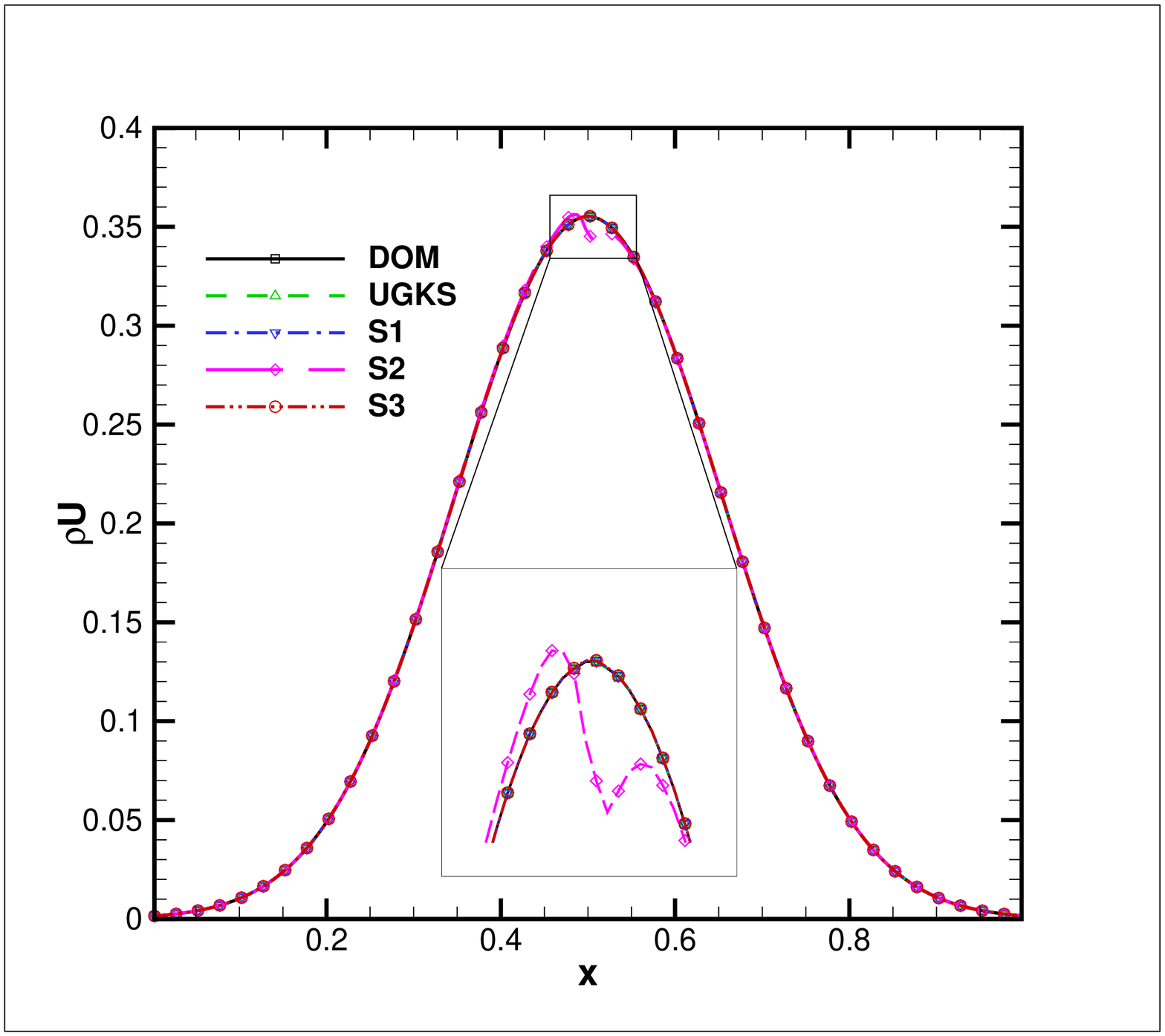}
    }
    \caption{The density and momentum profile of the shock tube problem at Kn$=10$.}
    \label{fig:sod2}
\end{figure}

\section{Numerical comparison}
In all the following numerical tests, the CFL number is 0.4. And all the numerical setting are exactly identical except the numerical flux for different numerical schemes.
\subsection{Sod shock tube}
At first, the one dimensional shock tube problem is tested under different Knudsen numbers.
\begin{eqnarray}
  \mbox{Kn} = \frac{\mu_{ref}\sqrt{RT_{ref}}}{p_{ref}L}. \label{eq:Kn}
\end{eqnarray}
The computational domain is $[0,1]$ in $x$ direction. And it is discretized into 200 cells. The initial condition is given as follows,
\begin{eqnarray}
  \left\{
  \begin{array}{llll}
    \rho_l = 1.0,& U_l = 0.0,& p_l = 1.0,&\quad \mbox{for}\  x \leq 0.5, \\
    \rho_r = 0.125,& U_r = 0.0,& p_r = 0.1,&\quad \mbox{for}\  x > 0.5.
  \end{array}\right.
\end{eqnarray}
The quantities on the right-half domain are selected to define the Knudsen number.
We use 150-point uniform grid in the velocity space $[-6,6]$. The computation stops at $t = 0.15$.
Figure \ref{fig:sod1} and \ref{fig:sod2} show the numerical results for Kn$=0.0001, 0.01, 1, 10$. Five different flux solvers are employed to simulate this problem. As expected, all the methods provide very good results.

In the free molecular flow regime, say, the Knudsen number is 10, we find that, except the S2 scheme, all the numerical schemes predict the same density and momentum profile in figure \ref{fig:sod1}. This is because the leading order terms are identical for all these schemes (Tab.(\ref{tab:coe}), Eq.(\ref{eq:domflux},\ref{eq:ugksflux},\ref{eq:s1flux})).
%\begin{eqnarray}
%  \mathcal{F}^W &=& \Delta t\sum_k u_k \psi_k f_k - \frac{1}{2}\Delta t^2 \sum_k u_k \psi_k\mathbf{u}_k\cdot\nabla f_k, \nonumber\\
%  \mathcal{F}_k &=& \Delta t u_k f_k - \frac{1}{2}\Delta t^2 u_k \mathbf{u}_k\cdot\nabla f_k.
%\end{eqnarray}
Inaccurate results from the S2 scheme verify that, a simple combination of the DOM and a Navier-Stokes solver cannot lead to correct asymptotic limit.
The quantities plotted in figure \ref{fig:sod1} are the macroscopic variables updated by the Eq.(\ref{eq:macroConserve}). When the relaxation time $\tau$ goes to infinity, the evolution of the distribution function (Eq.(\ref{eq:lastTimeStepCE})) are totally independent to the evolution of the macroscopic variables (Eq.(\ref{eq:macroConserve})), since the collision term vanishes. As a result, though the macroscopic variables are incorrect in the S2 results, the distribution function derived in the same simulation is identical to the other methods. We will demonstrate it in next two dimensional simulation.
In the transition flow regime, the results derived from different schemes are still indistinguishable. In the continuum flow regime, the S1, S2, S3, DOM and UGKS provide almost identical solution. It testified that, the inaccuracy of the initial distribution function affects little to the numerical performance in the continuum flow regime.

These numerical observation are consist with our analysis in the previous section. The discrepancy is hardly noticed in all the flow regimes. All the numerical methods (except the S2) converge to the Euler solution in the continuum regime, and converge to collisionless solution in the free molecular flow regime.
%\addtocounter{figure}{-1}
\subsection{Lid-driven cavity flow}
The one dimensional numerical results show that all the numerical schemes converge to the Euler solution at Kn$\rightarrow 0$. However, as mentioned in the reference \cite{chen2015comparative}, the one dimensional numerical experiment cannot distinguish the NS AP scheme from the Euler AP scheme. Thus we simulate a two dimensional lid-driven cavity flow which is characterized by strong viscous effect. The gas flow is confined in a square domain whose extent is $[0,1]\times[0,1]$. Each edge of the computational domain is uniformly discretized by 61 nodes. The top boundary moves from left to right with a constant velocity, 0.2. The gas pressure is 1; and the density is also 1. The Mach number based on the velocity of the top wall is about 0.15.
The Knudsen number is defined as Eq.(\ref{eq:Kn}).

\begin{figure}
    \centering
    \mbox{Kn=2.0, Re=0.1}
    \parbox[b]{0.93\textwidth}{
        \parbox[b]{0.45\textwidth}{
            \centering
            \mbox{(a)}
            \includegraphics[totalheight=6.5cm, bb = 92 42 680 545, clip=true]{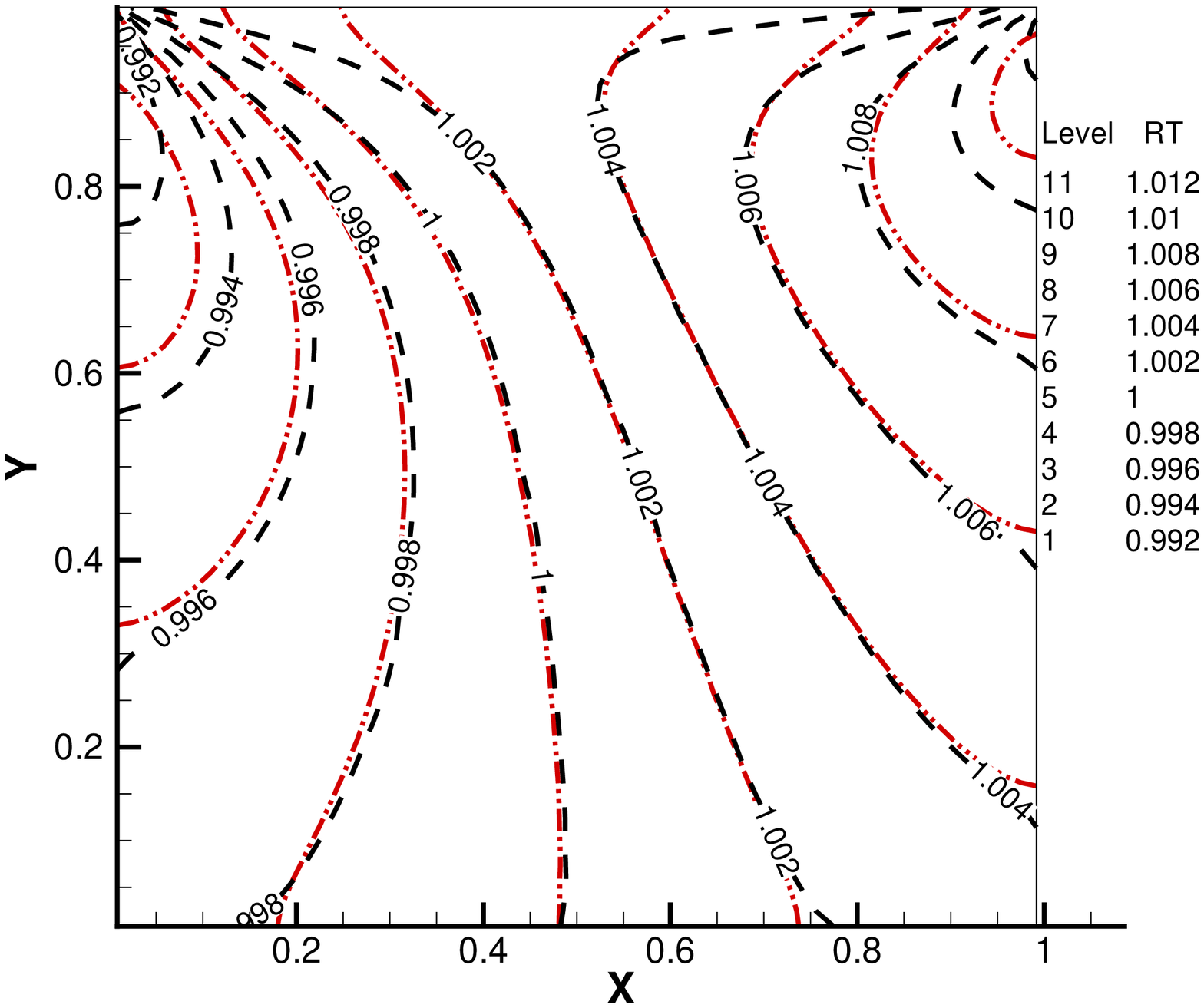}
        }\hfill
        \parbox[b]{0.45\textwidth}{
            \centering
            \mbox{(b)}
            \includegraphics[totalheight=6.5cm, bb = 92 42 680 545, clip=true]{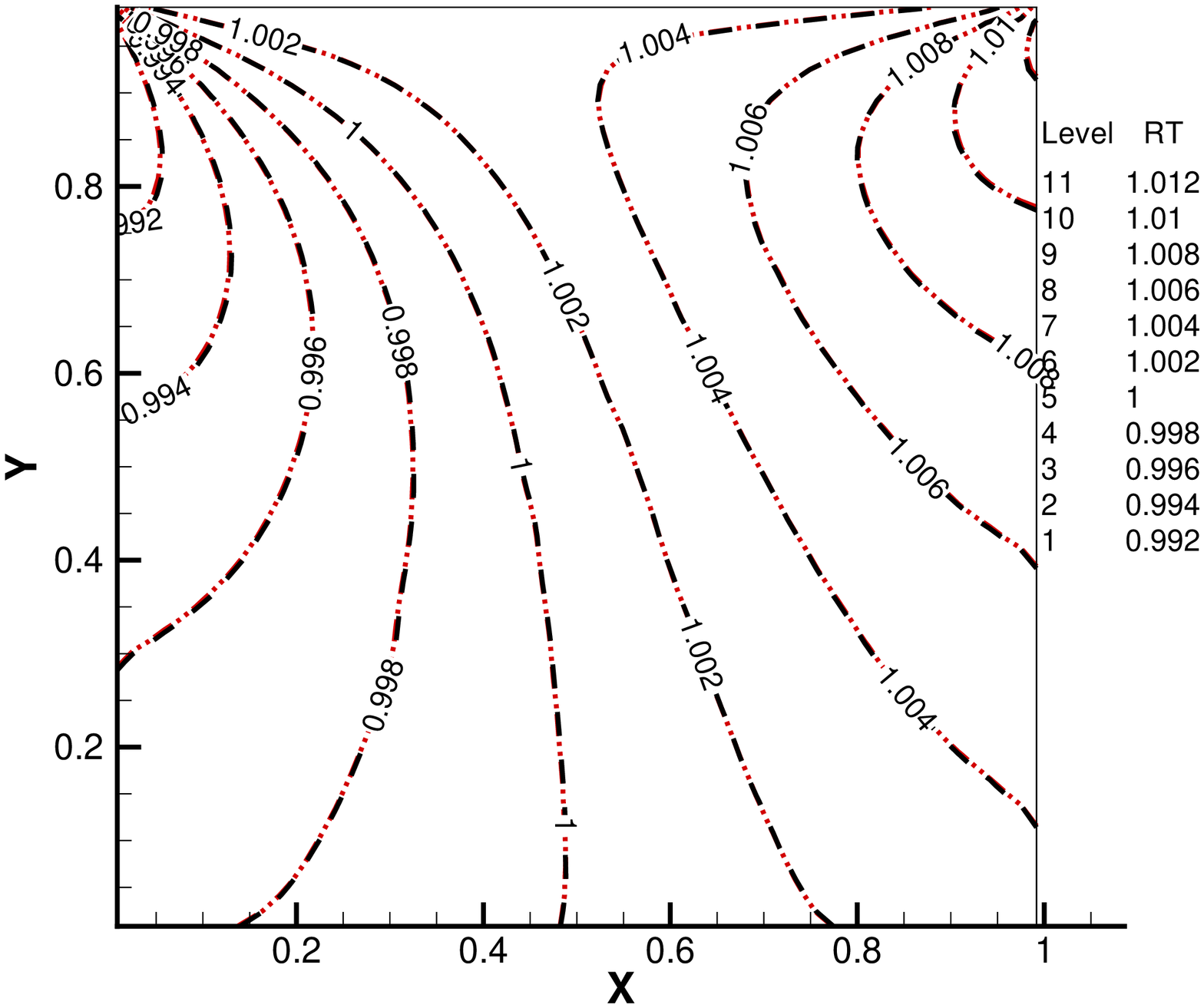}
        }
    }
    \caption{The temperature ($RT$) contours in the lid-driven cavity flow at Kn$=2$, Re$=0.1$. The black dash lines represent the $f$-based temperature from the DOM. (a) The red dash dot lines represent the $W$-based temperature from the S2 scheme; (b) The red dash dot lines represent the $f$-based temperature from the S2 scheme.}
    \label{fig:cavity1}
\end{figure}

\begin{figure}
    \centering
    \mbox{Kn=2.0, Re=0.1}
    \parbox[b]{0.95\textwidth}{
        \parbox[b]{0.45\textwidth}{
            \centering
            \mbox{(a)}
            \includegraphics[totalheight=6.5cm, bb = 92 42 680 545, clip=true]{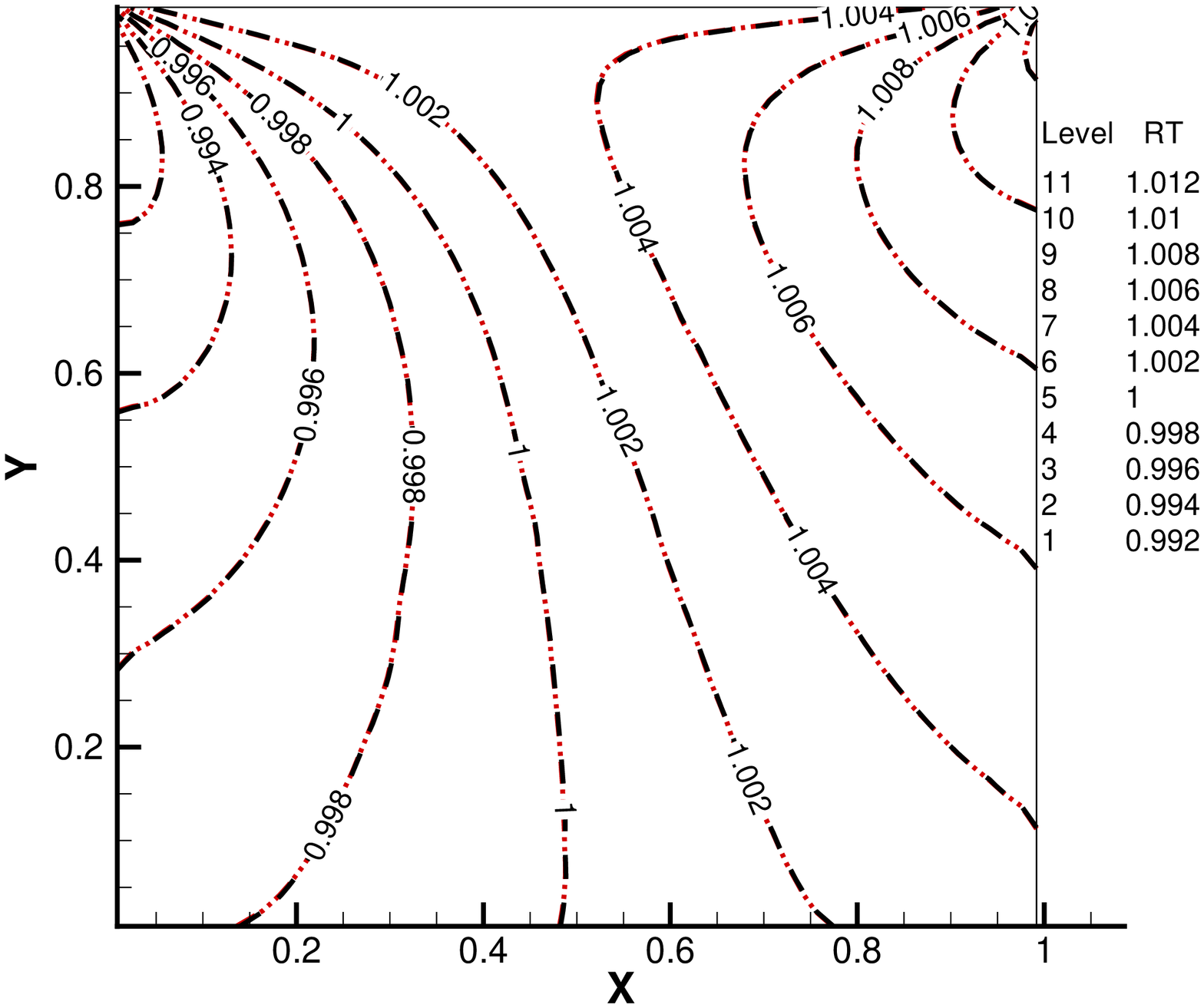}
        }\hfill
        \parbox[b]{0.45\textwidth}{
            \centering
            \mbox{(b)}
            \includegraphics[totalheight=6.5cm, bb = 92 42 680 545, clip=true]{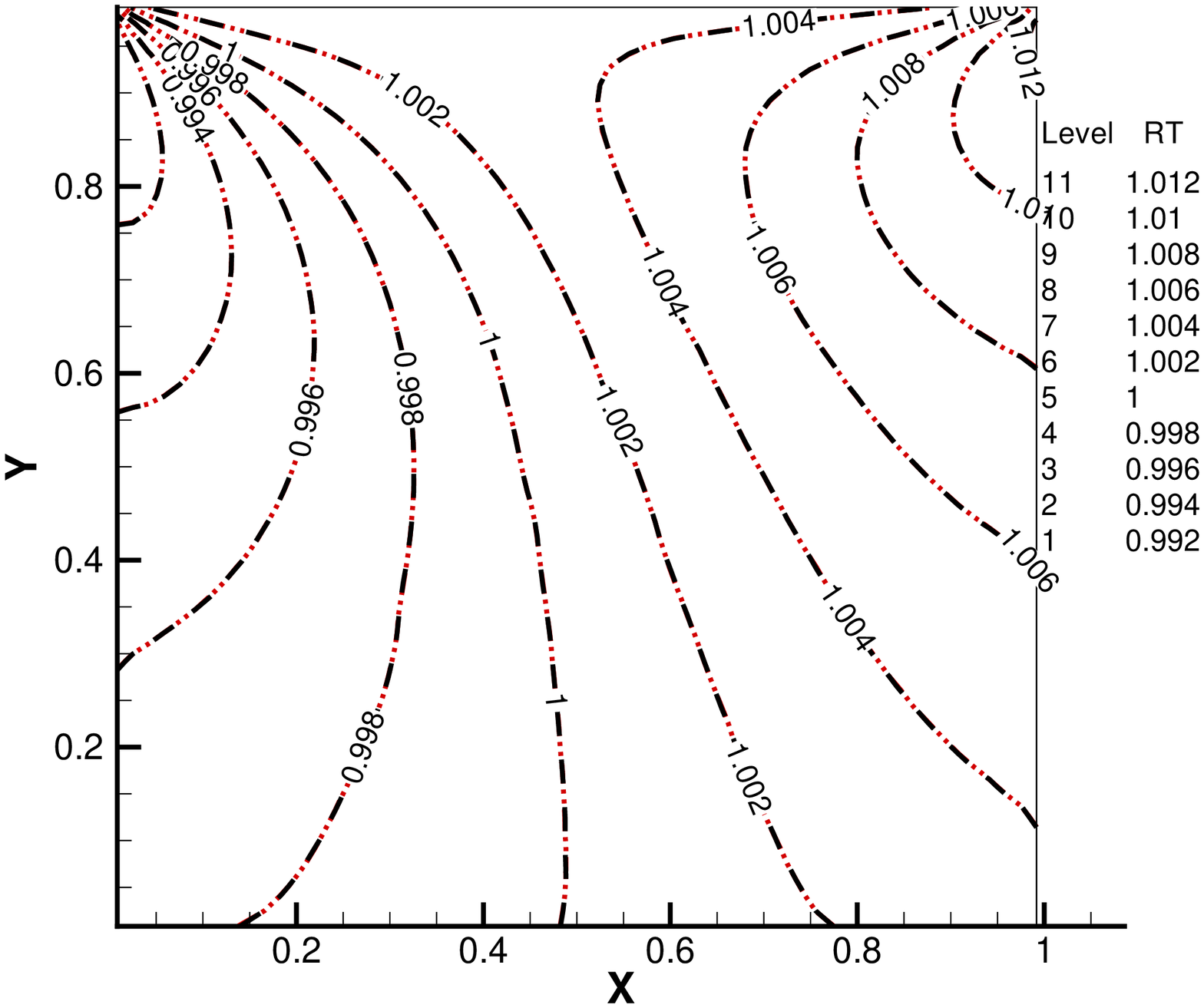}
        }
    }
    \caption{The temperature ($RT$) contours in the lid-driven cavity flow at Kn$=2$, Re$=0.1$. The black dash lines represent the $f$-based temperature from the DOM. (a) The red dash dot lines represent the $W$-based temperature from the S3 scheme; (b) The red dash dot lines represent the $f$-based temperature from the S3 scheme.}
    \label{fig:cavity2}
\end{figure}

\begin{figure}
    \centering
    \mbox{Kn=2.0, Re=0.1}
    \parbox[b]{0.95\textwidth}{
        \centering
        \includegraphics[totalheight=6.5cm, bb = 92 42 680 545, clip=true]{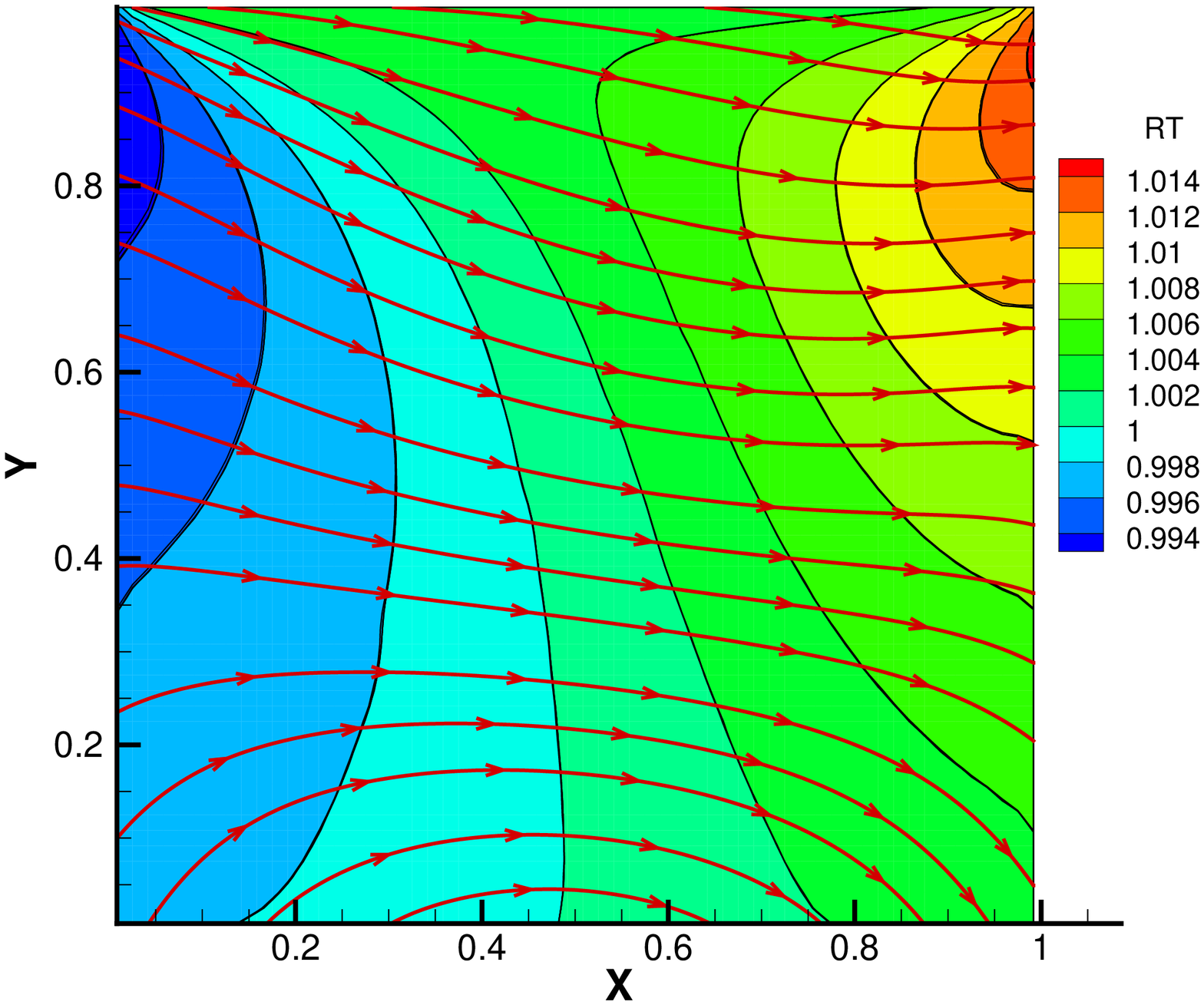}
    }
    \caption{The $f$-based temperature ($RT$) contour and the $f$-based heat flux in the lid-driven cavity flow at Kn$=2$, Re$=0.1$ for all the numerical schemes (include DOM, UGKS, S1, S2, and S3). The numerical results derived from different numerical schemes collapse to the DOM results.}
    \label{fig:cavity3}
\end{figure}

As mentioned in the last subsection, the conservative variables $W$ and the distribution function $f$ are evolving separately in the free molecular flow regime. Therefore, we use the $W$-based variable to denote the macroscopic variable deduced from the conservative variables $W$, and use $f$-based variable to denote the macroscopic variable deduced from the distribution function $f$. Figure \ref{fig:cavity1} shows the $W$-based temperature and the $f$-based temperature derived from the S2 scheme. The flow condition is $\mbox{Kn}=2$ and $\mbox{Re} = 0.1$, and the velocity space $[-5,5]\times[-5,5]$ is discretized into $100\times100$. Since the DOM is accurate at high Knudsen number, we choose the $f$-based temperature derived from the DOM as benchmark solution, and plot it on the background. As shown in figure \ref{fig:cavity1}, the $W$-based temperature deviates from the DOM solution. Meanwhile, the $f$-based temperature is identical to the DOM solution.
This is because that, $\mathcal{F}^{W}_{s2} \neq \langle \mathcal{F}_{s2} \rangle$ when $\beta$ approaches zero, namely, the macroscopic flux is inconsistent with the flux of distribution function. More specifically, this is the immediate consequence of the incorrect asymptotic coefficient $\gamma_3^{s2}$ in the S2 scheme (Tab.(\ref{tab:coe})). After remedying the coefficient, the S3 scheme has the same asymptotic limit as the analytical solution. As we can see in figure \ref{fig:cavity2}, the results obtained from the S3 scheme, both $f$-based and $W$-based temperatures coincide with the results derived from the DOM. The results from all the considered numerical methods collapse to the DOM results in figure \ref{fig:cavity3}.

\begin{figure}
    \centering
    \mbox{Kn=0.0002, Re=1000}
    \parbox[b]{0.95\textwidth}{
        \parbox[b]{0.45\textwidth}{
            \centering
            \mbox{(a)}
            \includegraphics[totalheight=6.5cm, bb = 92 25 690 545, clip=true]{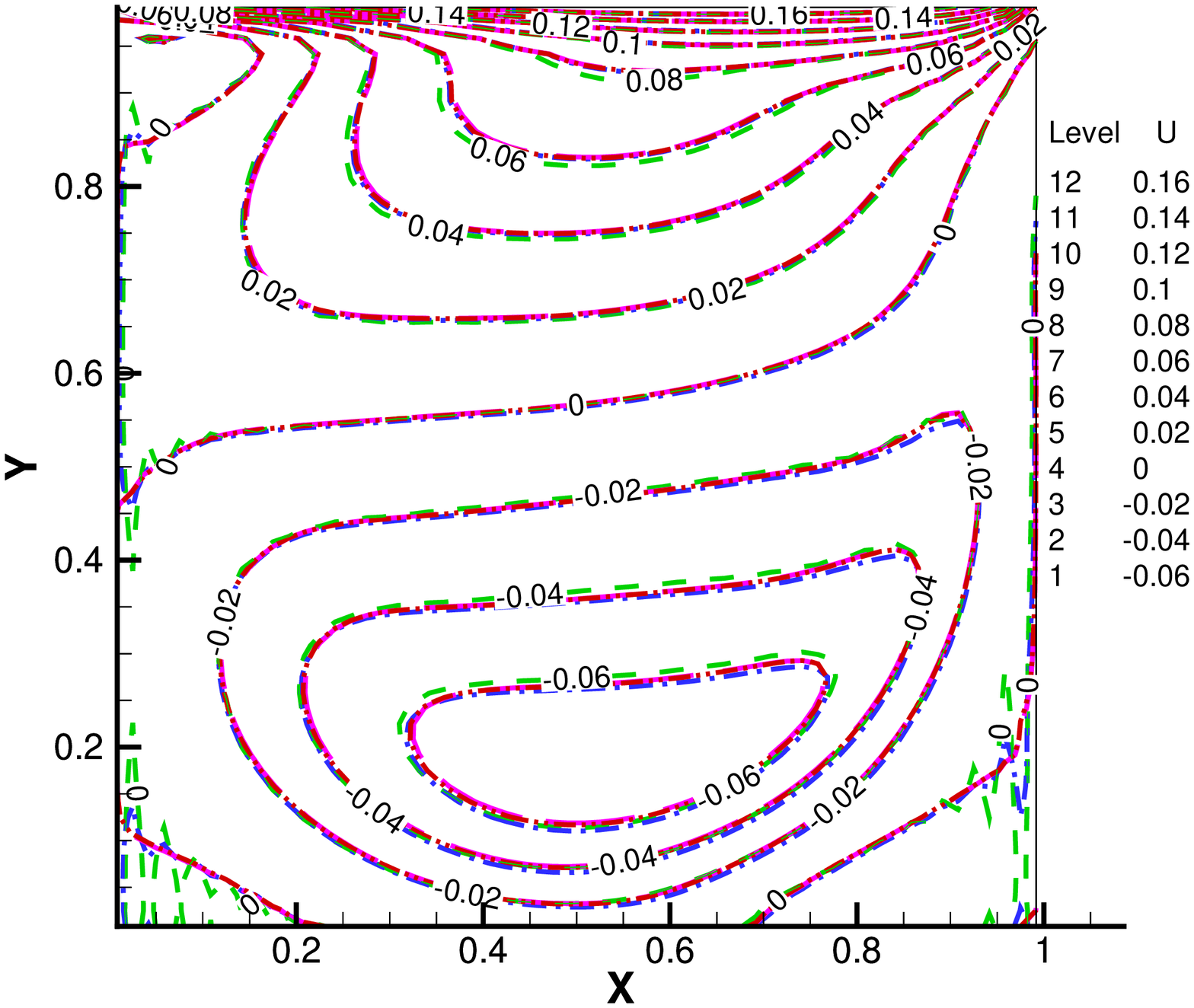}
        }\hfill
        \parbox[b]{0.43\textwidth}{
            \centering
            \mbox{(b)}
            \includegraphics[totalheight=6.5cm, bb = 92 25 680 545, clip=true]{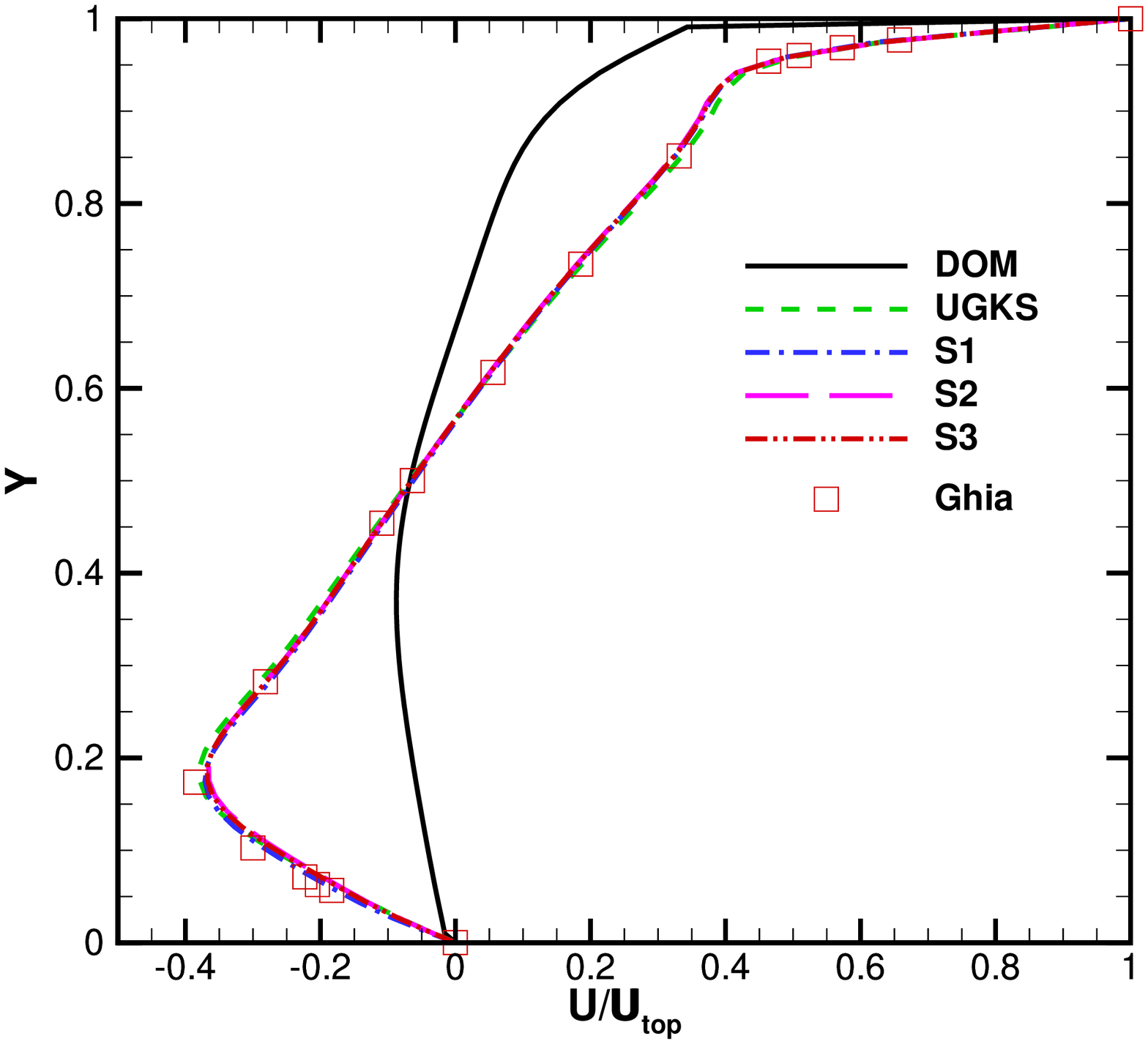}
        }
    }
    \parbox[b]{0.95\textwidth}{
        \parbox[b]{0.45\textwidth}{
            \centering
            \mbox{(c)}
            \includegraphics[totalheight=6.5cm, bb = 92 25 690 545, clip=true]{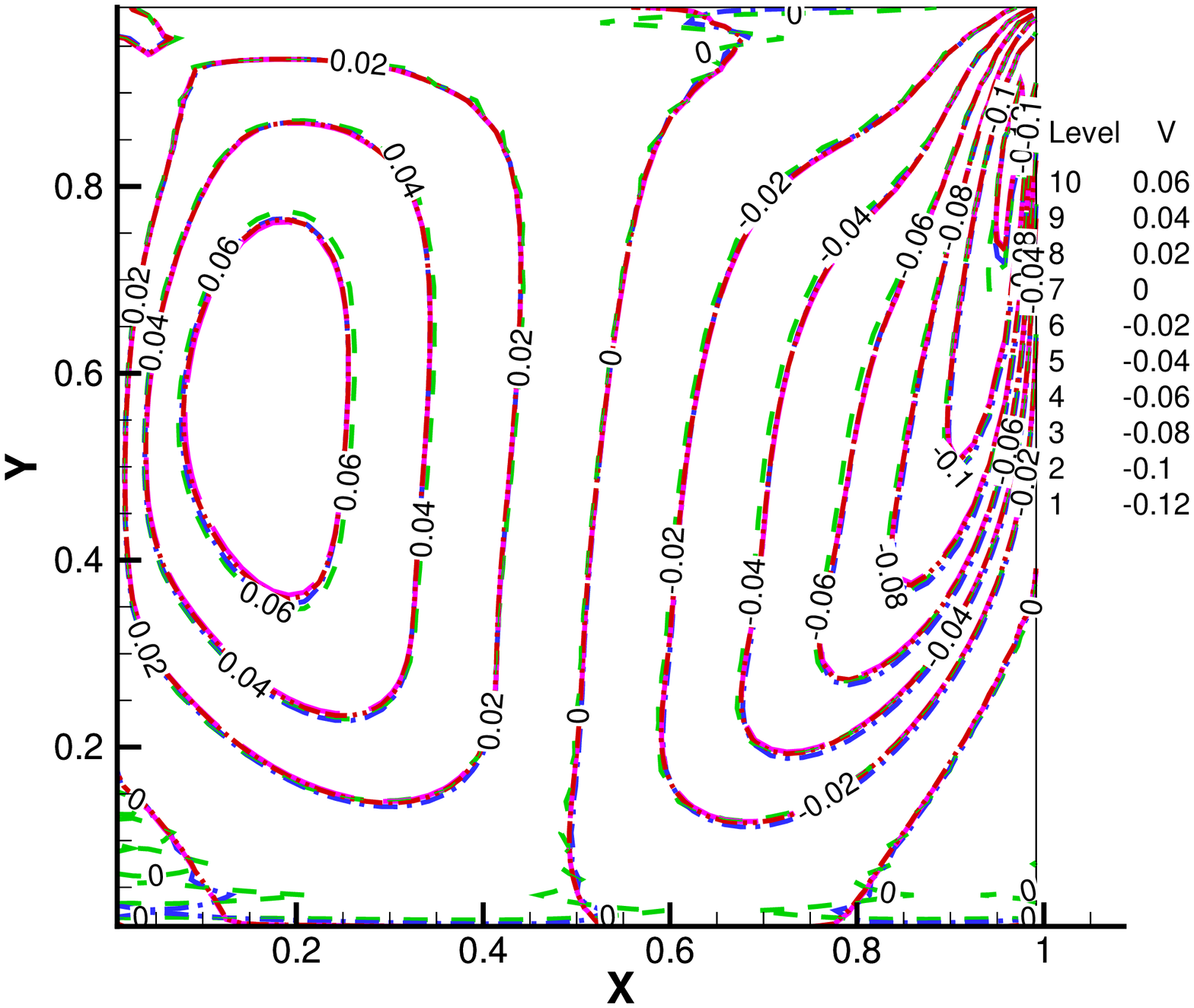}
        }\hfill
        \parbox[b]{0.43\textwidth}{
            \centering
            \mbox{(d)}
            \includegraphics[totalheight=6.5cm, bb = 92 25 680 545, clip=true]{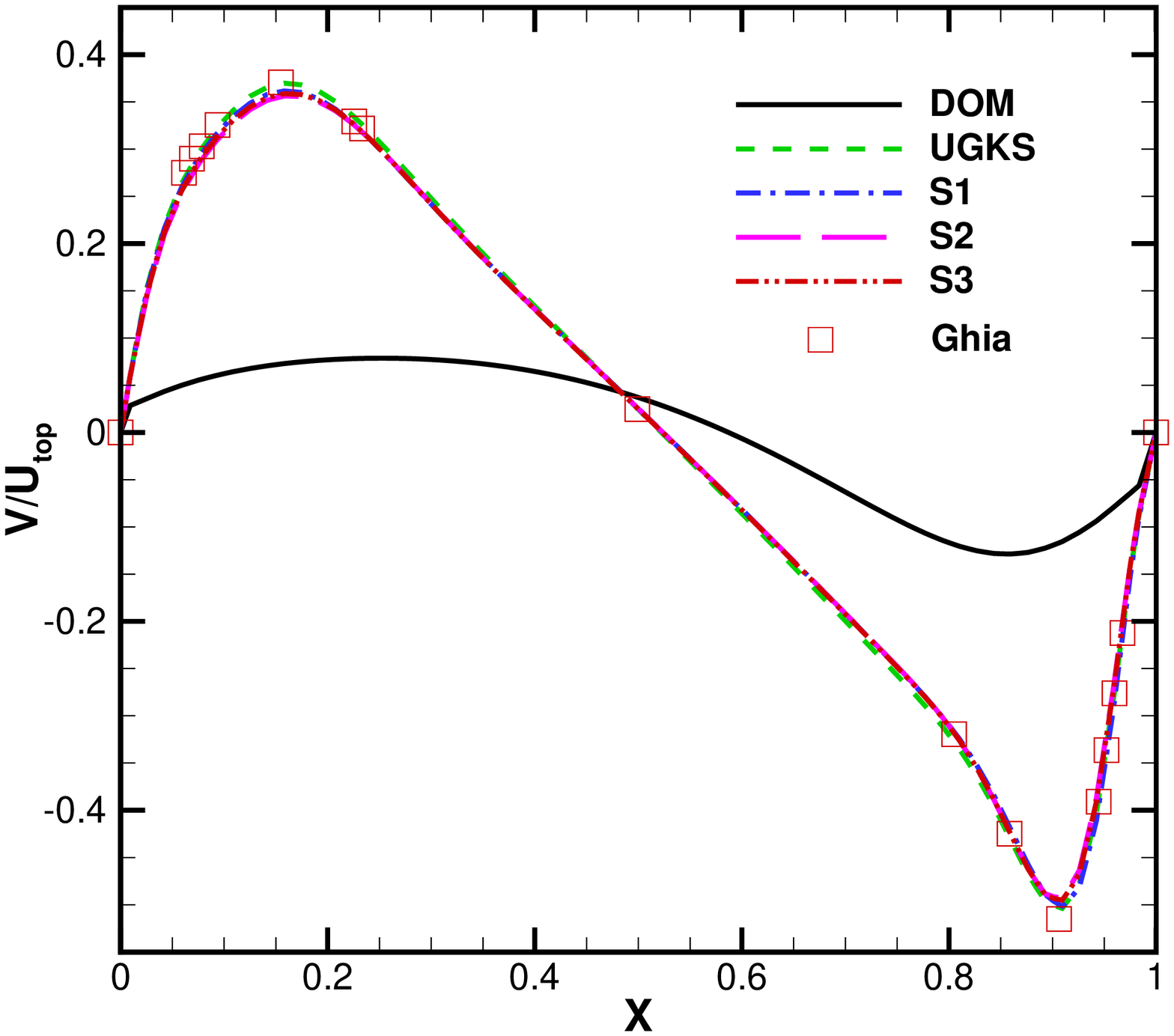}
        }
    }
    \caption{The velocity contours (a, c) and velocity profiles (b, d) at Kn = 0.0002, Re = 1000.}
    \label{fig:cavity4}
\end{figure}

But remarkable discrepancies are observed when the Reynolds number increases to 1000.
In this case, we only use 8 velocity points in one direction to discretize the velocity space ranging from -5 to 5. And the rectangular quadrature in velocity space is adopted. All these numerical settings are on the purpose of illustrating the influence of the inaccurate quadrature in velocity space. Central difference interpolation is adopted for both microscopic and macroscopic variables.
As shown in Fig. \ref{fig:cavity4}(b,d), the DOM cannot simulate the continuum flow properly, therefore, the DOM's results are not shown in Fig. \ref{fig:cavity4}(a,c). The UGKS and S1 schemes obtained much better numerical results which are closer to the reference data \cite{Ghia1982}. However, due to the inaccuracy of the quadrature in the velocity space, the numerical results are not as good as the numerical results in the previous literatures\cite{ugks2,wang2015comparative,zhu2015performance}. The numerical contour lines oscillate near the boundaries (Fig. \ref{fig:cavity4}(a,c)). On the other hand, the simplified schemes (S2, S3) perform best in this test case.

\begin{figure}
    \centering
    \mbox{Kn=0.02, Re=10}
    \parbox[b]{0.95\textwidth}{
        \parbox[b]{0.45\textwidth}{
            \centering
            \mbox{(a)}
            \includegraphics[totalheight=6.5cm, bb = 92 25 690 545, clip=true]{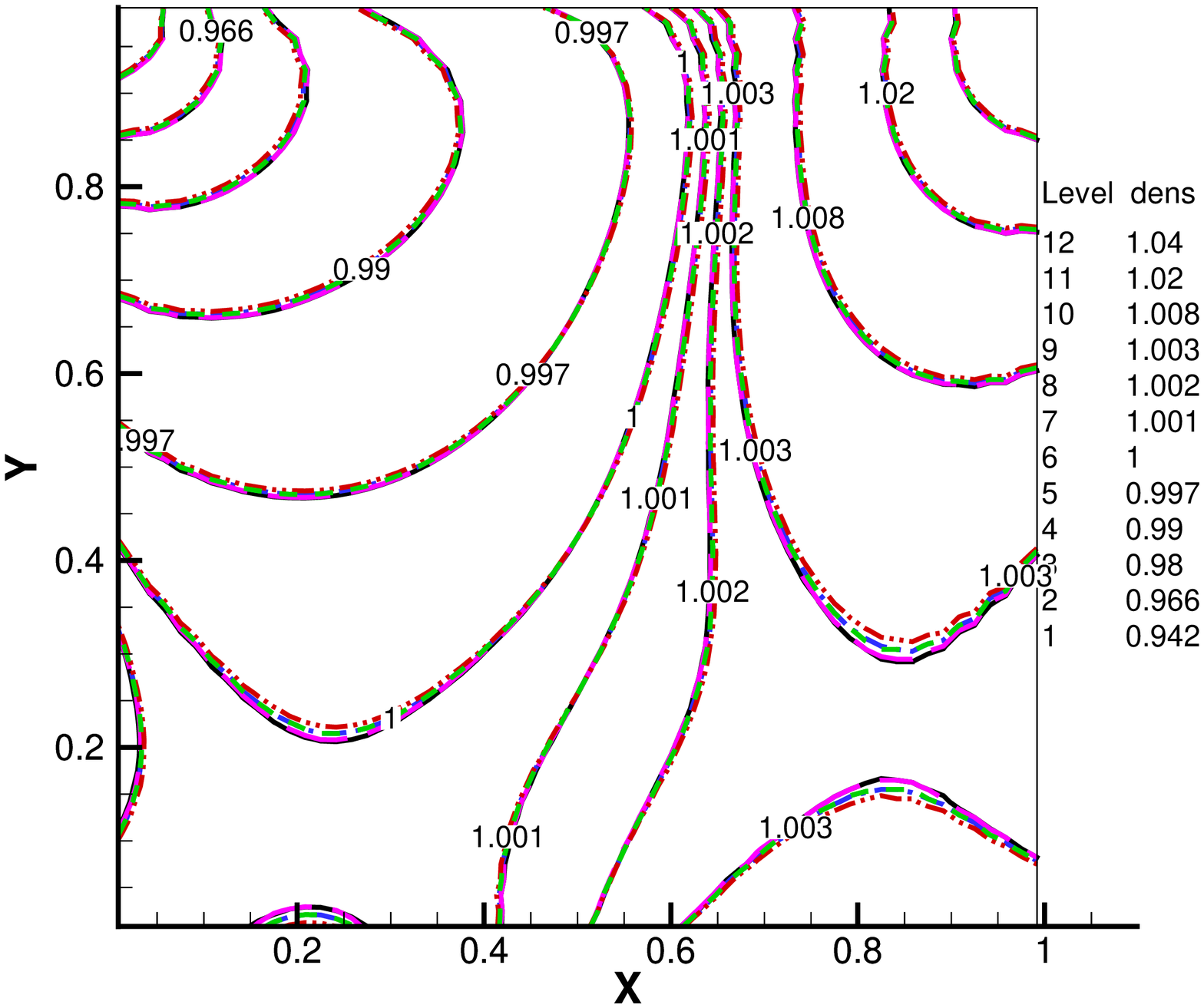}
        }\hfill
        \parbox[b]{0.45\textwidth}{
            \centering
            \mbox{(b)}
            \includegraphics[totalheight=6.5cm, bb = 92 25 700 545, clip=true]{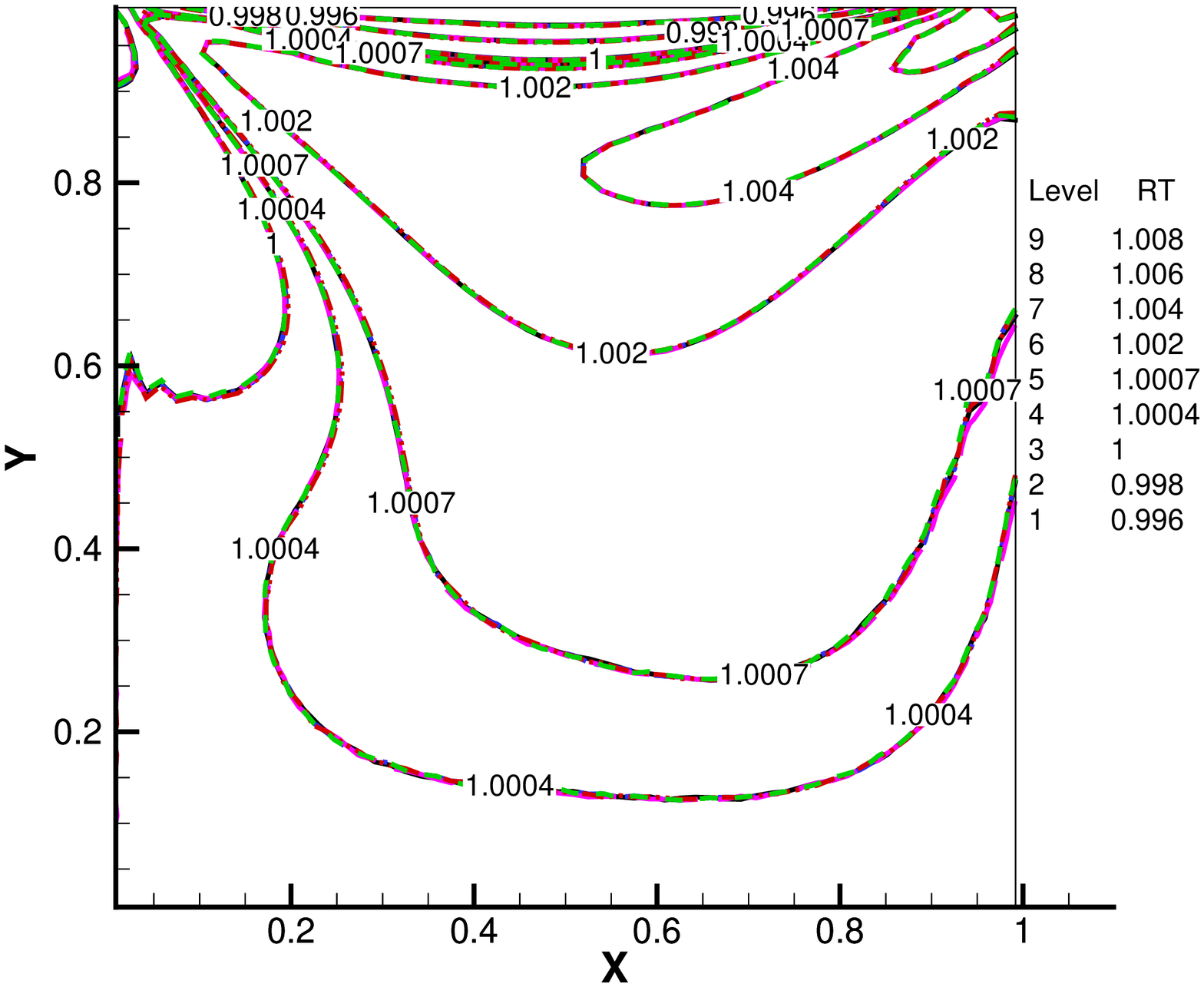}
        }
    }
    \parbox[b]{0.95\textwidth}{
        \parbox[b]{0.45\textwidth}{
            \centering
            \mbox{(c)}
            \includegraphics[totalheight=6.5cm, bb = 92 25 690 545, clip=true]{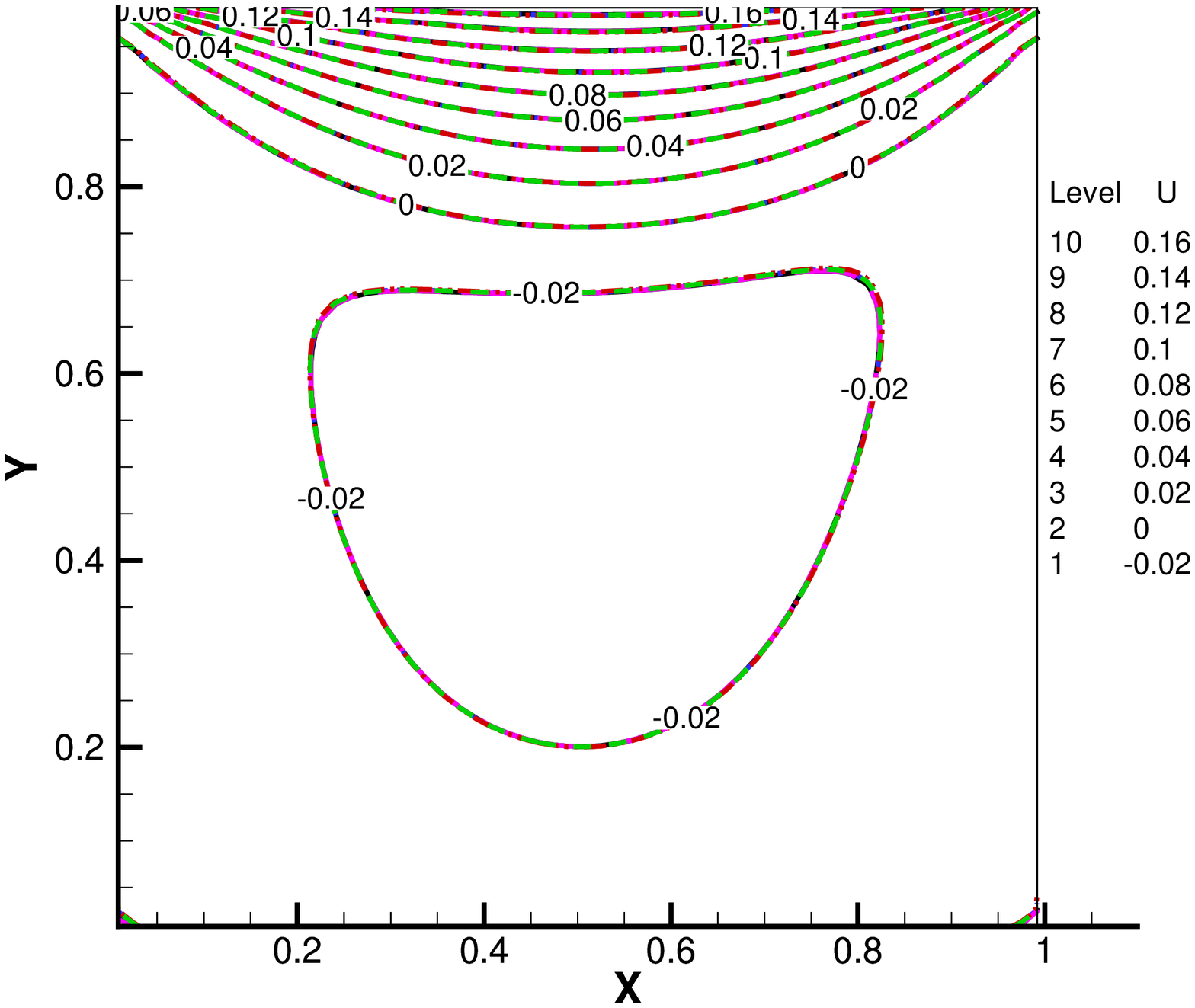}
        }\hfill
        \parbox[b]{0.45\textwidth}{
            \centering
            \mbox{(d)}
            \includegraphics[totalheight=6.5cm, bb = 92 25 680 545, clip=true]{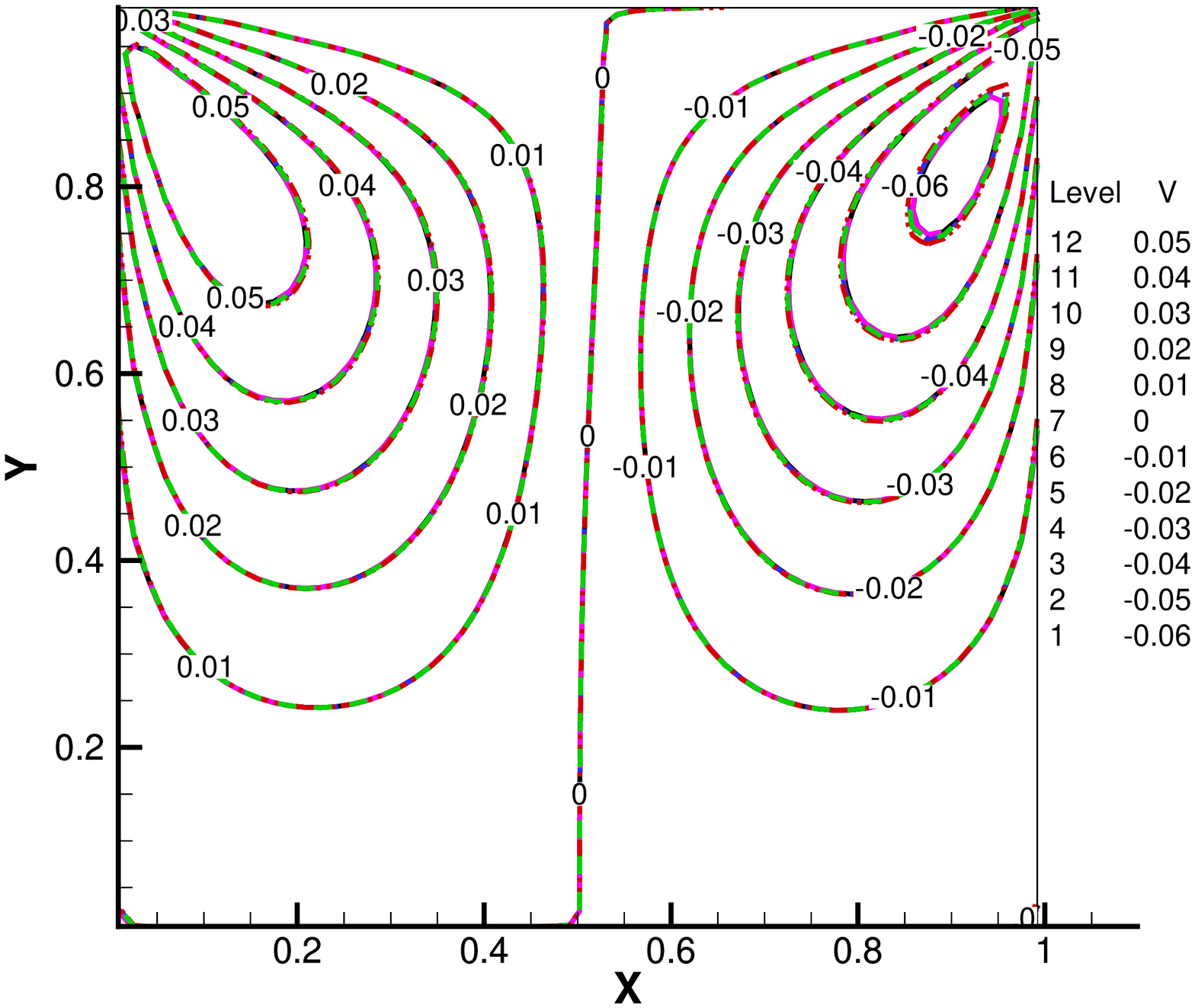}
        }
    }
    \caption{The flow field at Kn = 0.02, Re = 10. (a) The density contour; (b) The temperature contour; (c,d) The velocity contour. DOM: black solid line; UGKS: green dash line; S1: blue dash dot line; S2: pink long dash line; S3: red dash double dot line.}
    \label{fig:cavity5}
\end{figure}

The asymptotic limits of the numerical schemes coincide with our analysis in the previous section. For the transition flow regime, the numerical results are shown in figure \ref{fig:cavity5}. The Reynolds number is 10, and the Knudsen number is 0.02. We use 61 points in physical space, and use 60 points in velocity space.
The velocity contours are almost identical for all the schemes. Only minor differences can be noticed in density contour and temperature contour.

From the above results, we demonstrate that the simplified schemes proposed in the paper possess correct asymptotic limit in free molecular flow regime and the continuum flow regime, and provide enough accurate numerical results in transition flow regime.

%\subsection{2D supersonic flow past a cylinder}
\subsection{The high efficiency of the simplified methods}
In the Eq.(\ref{eq:s1flux}), three out of five terms are evaluated by analytical formulas. These computational costs are infinitesimal compared to the quadrature in velocity space. We also observe that the S1 reduces about half computation time compared to the UGKS. And the DOM, S1, S2, and S3 schemes have almost identical computational efficiency.

On the other hand, the numerical results derived from the simplified methods are closer to the results from NS solver in the continuum flow regime. It is worth noting that, the coefficient for S2, say, $\gamma_0^{s2}$, deviates from $0$ by a exponential truncation error, while $\gamma_0^{ugks}$ and $\gamma_0^{s1}$ preserve $\tau$ as the leading order term. As illustrated in Eq.(\ref{eq:ugksfluxW},\ref{eq:assumptionCE},\ref{eq:ugksCE}), the physical asymptotic process is not simply attained by vanishing the non-equilibrium terms, $f_0$ and $\mathbf{u}\cdot \nabla f$. The non-equilibrium terms still contribute a little $(O(\tau))$ to the total distribution function, and the remaining terms of non-equilibrium part are canceled by the equilibrium part, then result in the Chapman-Enskog expansion. Such balance is very delicate and sophisticated. It is definitely computationally burdensome or clumsy to simulate this subtle asymptotic process in velocity space. The simplified methods proposed in this study circumvent the delicate balance, instead, use more rapid decaying coefficients in front of the non-equilibrium terms. The quadrature of $f_0$ in velocity space impose almost nothing on the numerical macroscopic flux which means less numerical error in the scheme.
As we can see in the numerical comparisons, the S2 and S3 schemes provide more accurate numerical results in the continuum flow regime, since the delicate balance between the non-equilibrium part and the equilibrium part are replace by a prior knowledge and circumvent the numerical simulation of asymptotic process. The quadrature of the distribution function is totally replaced by the analytical expression. Hence the simplified schemes lead to more accurate results, and less discrete points in velocity space.

\section{Conclusion}
In this study, we analyzed the asymptotic behavior of the unified gas kinetic scheme, and reduced the unnecessary quadrature in the UGKS numerical flux for the equilibrium part. In the first simplified scheme, the quadrature in velocity space for the equilibrium part is replaced by the analytical results. The numerical comparison shows that this replacement reduces about half computation load and does not effect numerical results. Based on the asymptotic expression of the coefficients in the UGKS flux, several other simplification strategies have been proposed. The numerical comparisons demonstrated that simple combination (S2) of a kinetic flux and the macroscopic flux cannot obtain correct asymptotic limit in the free molecular flow regime. With a rescaled viscosity coefficient, the simplified scheme (S3) possesses correct asymptotic limit both in the free molecular flow regime and in the continuum flow regime. Moreover, it can be constructed by combining two existing flux solvers which handle the kinetic equation and the Navier-Stokes equations respectively. The simplified scheme (S3) is efficient in terms of coding and computing, hence, is a promising approach for engineering application. Its accuracy is also acceptable and controllable. The flux hybrid strategy proposed in this study can be further extended to the other multiscale problems.

\section*{Acknowledgements}
This work was supported by NSF91530319, Hong Kong Research Grant Council (620813,
16211014,
16207715), and HKUST (PROVOST13SC01,
IRS15SC29,
SBI14SC11).

\section*{References}

%\bibliography{SUGKS_arxiv}
%\bibliography{../CartesianMesh}

%

%\nocite{*}

\end{document}